\documentclass[preprint,3p]{elsarticle}

\journal{}
\usepackage{graphicx}% Include figure files
\usepackage{dcolumn}% Align table columns on decimal point
\usepackage{bm}% bold math
\usepackage{bigints}
\usepackage[mathlines]{lineno}% Enable numbering of text and display math
\usepackage{comment}
\linenumbers\relax % Commence numbering lines

\usepackage[utf8]{inputenc}
\usepackage[T1]{fontenc}
\usepackage{amsmath,amssymb}
\usepackage{etoolbox}
\usepackage{bbold}
\usepackage{color}
\usepackage{pgfplots}
\pgfplotsset{compat=newest}
\usepackage{tikz}
\usetikzlibrary{patterns}
\newcommand{\opt}{\mathrm{opt}}

\makeatletter
\def\thickhline{%
  \noalign{\ifnum0=`}\fi\hrule \@height \thickarrayrulewidth \futurelet
   \reserved@a\@xthickhline}
\def\@xthickhline{\ifx\reserved@a\thickhline
               \vskip\doublerulesep
               \vskip-\thickarrayrulewidth
             \fi
      \ifnum0=`{\fi}}
\makeatother

\newlength{\thickarrayrulewidth}
\journal{Journal of Sound and Vibration}

\begin{document}
\nolinenumbers

\begin{frontmatter}

\title{Optimal damping adapted to a set of initial conditions}
%or Optimal damping adapted to a given set of initial conditions
\author[1]{K. Lelas\corref{cor1}}
\ead{klelas@ttf.unizg.hr}

%\author[2]{I. Nakić}
%\ead{nakic@math.hr}

\affiliation[1]{
organization={Faculty of Textile Technology, University of Zagreb},
country={Croatia}
}
%\affiliation[2]{
%organization={Department of Mathematics, Faculty of Science, University of Zagreb},
%country={Croatia}
%}
\cortext[cor1]{Corresponding author}


\begin{abstract}
Vibrating systems can respond to an infinite number of initial conditions and the overall dynamics of the system can be strongly affected by them. %For real systems in operating conditions, the excitations can be such that some initial conditions are often realized, and some rarely or not at all.
Therefore, it is of practical importance %,posebno za sisteme s passivnim prigušenjem
to have methods by which we can determine the damping that is in some sense optimal for all initial conditions, or for a given set of initial conditions. For a single and multi degree of freedom systems, we determine the optimal damping coefficients adapted to different sets of initial conditions using the known method of minimizing the (zero to infinity) time integral of the energy of the system, averaged over a set of initial conditions, and using two new methods that we introduce. One method is based on determining the damping for which the energy of the system, averaged over a set of initial conditions, drops the fastest to a given threshold value. The other method is based on determining the damping that gives minimal average settling time of the system, where we take that the system settled when its energy dropped to a given threshold value. We show that the two new methods give results for optimal damping that are in excellent agreement with each other, but are significantly different from the results given by the minimization of the average energy integral. More precisely, for considered multi degree of freedom systems and sets of initial conditions, the two new methods give optimal damping coefficients that converge to the critical damping of the first mode as the target energy threshold decreases. On the other hand, for these same systems and sets of initial conditions, the method of minimizing the average energy integral gives optimal damping coefficients which are deep in the overdamped regime with respect to the first mode.  
\end{abstract}


\end{frontmatter}

\section{Introduction}

In most methods for determining the optimal damping of multi-degree of freedom (MDOF) linear vibrating systems initial conditions are either completely ignored or zero initial conditions are assumed \cite{rojas2019optimal, fu2001modal, takewaki1997optimal, CHEUNG200929, ALUJEVIC20144073}. The reason for this is that these methods are primarily designed for the analysis of systems with continuous excitation and finding the optimal steady state response, and are not intended for the analysis of free vibrations with non-zero initial conditions. In some cases it make sense to study the free vibrations of the system with nonzero initial conditions.  %ako recezent pita zasto nisu suitable, onda opis tih metoda, frekvencijska domena itd, sad mi se ne da...iz proslog clanka: Specifically, for methods based on frequency domain approach, which are designed for forced vibrations, it is not clear how to take into account the non-zero initial conditions in a systematic way.  
A prominent example is the vibration control in buildings exposed to seismic excitations \cite{ventura1992influence, wang2007vibration}. Indeed, depending on the initial conditions, MDOF systems can exhibit oscillatory or non-oscillatory response \cite{morzfeld2013characterization}, so it is clear that initial conditions can play an important role in the overall dynamics of the system. Implicitly, dependence of the behavior of system on initial conditions has been investigated in the context of time-optimal vibrations reduction \cite{dhanda2016vibration} and transient response \cite{meirelles2005transient} in terms of computationally efficient methods for the calculation of the system response.%kao primjeri slobodnih vibracija i poc uvjeta mogu se navesti jos i residualne vibracije kod uredjaja, shock inducirane vibracije, imapact inducirane vibracije itd itd

In \cite{Ves90}, optimal damping of free vibrations was determined by minimization of the (zero to infinity) time integral of the energy of the system, averaged over all possible initial conditions corresponding to the same initial energy. This method was investigated widely, mostly by mathematicians, in the last two decades, more details can be found, e.g., in references \cite{cox2004lyapunov, NakicPhd, truhar2009efficient, veselic2011damped, jakovvcevic2022fast}. %We will refer to this damping optimization method as \emph{the minimization of the average energy integral}. 
Due to the averaging over all initial conditions, it was not clear how well the optimal damping obtained in this way works for a specific choice of initial conditions, e.g., for free vibrations with initial conditions such that the initial energy consists only of potential energy, or only of kinetic energy, etc. This issue was recently addressed in \cite{Lelas2}, where a systematic investigation of the role of initial conditions in the context of optimal damping problems was initiated. Namely, optimal damping of free vibrations with specific initial conditions is determined simply by minimization of the energy integral corresponding to that particular initial conditions, i.e. without averaging, and it is found that optimal damping obtained this way strongly depends on initial conditions. In particular, for a single degree of freedom (SDOF) system, averaging over all initial conditions gives the critical damping as optimal \cite{Ves90,NakicPhd}, and by considering the minimization of the energy integral without averaging, damping coefficients approximately $30\%$ less than critical to infinite value are obtained as optimal, depending on the initial conditions \cite{Lelas2}. Similar behavior is found in MDOF systems with mass proportional damping (MPD). More precisely, for MDOF systems with MPD, minimization of the energy integral gives the lowest values of the optimal damping coefficients for initial conditions with purely potential initial energy, and infinite values of optimal damping coefficients for initial conditions with purely kinetic initial energy, while all values in between are possible for initial conditions with initial energy that is a combination of potential and kinetic energy. Furthermore, the optimal damping coefficients obtained in this way do not depend on the signs of the initial conditions, and the energy dissipation rate can strongly depend on them \cite{Lelas2}.

Along with the aforementioned analysis of optimal damping obtained by the minimization of energy integral for specific initial conditions, a new method for determining the optimal damping of MDOF systems was also introduced in \cite{Lelas2}. The method is based on the determination of the damping coefficients for which the energy of the system drops the fastest to some threshold that corresponds, e.g., to the energy resolution of the measuring device with which we observe the system, or to an acceptable level of energy that no longer poses a problem. This fastest energy drop method gives finite optimal damping coefficients even for initial conditions with purely kinetic initial energy, and it is sensitive to the signs of initial conditions as well. This method was introduced for the first time in \cite{Lelas}, but only in the context of the SDOF systems, where it was also experimentally verified (for one type of initial conditions) by measuring the optimal damping of free oscillations of the RLC circuit, i.e. measurements were consistent with theoretical predictions.    

Vibrating systems can respond to infinitely many different initial conditions in operating conditions. It begs the question, is it possible to adjust the damping to all initial conditions, or to some expected set of initial conditions? The method of minimizing the energy integral, averaged over initial conditions, is one way to determine the optimal damping adapted to all initial conditions, or some given set of initial conditions, the difference is only in the set over which we calculate the average. We will refer to this method as the \emph{minimization of the average energy integral}. The aim of this paper is to show that the fastest energy drop method, introduced in \cite{Lelas2}, can be adapted for this purpose as well. We consider the energy averaged over a set of initial conditions and take as optimal the damping for which this average energy drops to a given threshold the fastest. We will refer to this method as the \emph{fastest drop of the average energy}. For SDOF systems, and MDOF systems with MPD, we calculate and compare the optimal damping coefficients provided by those two methods for several sets of initial conditions. 

We find that these two methods give significantly different results when applied to the same set of initial conditions. Here we will highlight only a part of the results that perhaps most clearly indicate the differences between the two methods. For the set of all initial conditions with the same initial energy, we show that the fastest drop of the average energy gives optimal damping coefficients that converge to critical damping in case of SDOF, and to critical damping of the first modes in case of 2-DOF and 3-DOF systems with MPD, if we consider the energy drop to ever lower thresholds. For the same set of initial conditions, we show that the minimization of the average energy integral gives critical damping as optimal in the SDOF case, and it gives optimal damping coefficients that are significantly overdamped with the respect to the first modes in case of 2-DOF and 3-DOF systems with MPD. We argue that these differences in values of optimal damping coefficients given by two methods further increase with the increase of number of degrees of freedom.

If we would like to check which of the two methods gives more relevant results for real-world application, it would be best to do an experiment. Since this is a theoretical paper, we will only describe what kind of experiment could be useful for the selection of a "better" method. Let us imagine that we have an MDOF system and possibility to excite free vibrations of that system, so that each time some initial condition from a given finite discrete set of initial conditions (which correspond to the excitations to which the system will be exposed in operating conditions) is realized, and that each new free vibration starts after the previous one has ended. Also, let us imagine that we can adjust the damping in that system. In that case, we could set the damping to the value given by one method and, at least in principle, measure the duration of free vibrations corresponding to each initial condition from the given set. In this way, we could determine the total time of free vibrations of all initial conditions from the set, and calculate the average settling time of the system per initial condition. We could repeat the experiment with the optimal damping given by the other method and calculate the corresponding average settling time. We could then take the damping that gives a smaller average settling time of the system as relevant to the operating conditions.

This motivates us to introduce yet another method, the method of \emph{minimal average settling time}. %As we noted earlier, every measuring device has some resolution. Therefore%
We define the settling time of the system, corresponding to some specific initial condition, as a time needed for energy of the system to drop from initial value to some threshold value, and using the settling times corresponding to each initial condition in a given set, we calculate the average settling time of the system with respect to a given set of initial conditions. We repeat this procedure for a range of damping coefficients and search for the optimal damping coefficient for which the average settling time is minimal. We find that the results of this method are in excellent agreement with the results given by the fastest drop of the average energy. 

We focus mostly on vibrating systems with MPD so that we could analytically perform modal analysis and present new methods in the simplest possible way, but, we also show how to analyze optimal damping of vibrating systems with non-proportional damping. Using a 2-DOF system as an example, we show that the two new methods yield results for optimal damping that are in excellent agreement with each other in the case of non-proportional damping as well.  

The rest of the paper is organized as follows: Section 2 is devoted to SDOF systems, systematization of initial conditions and the sets of initial conditions that we consider in this paper, and precise definition and analysis of minimization of the average energy integral and fastest drop of the average energy methods in the SDOF case. In Section 3, we analyze MDOF systems with MPD. In Section 4, we define and analyze average settling time of SDOF and MDOF systems with MPD. In Section 5, we analyze optimal damping of a simple 2-DOF system with non-proportional damping using all three methods. In Sections 6 and 7, we discuss and summarize the important findings of the paper.
 
%Thus, we conclude that important theoretical, as well as practical, insights given by the minimization of the energy integral can be overlooked due to the averaging over all initial conditions....(ovdje reci da nasi rezultati pokazuju da, iako smo svjesni matematicke slozenosti tretiranja n dim sustava na ovaj nacin, nasi rezultati ukazuju da je to vrijedno istrazivanja u kontekstu primjenjivosti bla bla)  

\section{SDOF systems: minimization of the average energy integral and fastest drop of the average energy} 
\label{1Dsection}

\subsection{Short review of the model, systematization of initial conditions, average energy and average energy integral}
\label{1Dsubsec1}

Free vibrations of a SDOF linear vibrating system can be described by the equation 
\begin{equation}
\ddot q(t)+2\gamma\dot q(t)+\omega_0^2q(t)=0,\, q(0)=q_0, \, \dot{q}(0)=\dot{q}_0,
\label{DHOeq}
\end{equation}
where $q(t)$ denotes the displacement from the equilibrium position (set to $q=0$) as a function of time, the dots denote time derivatives, $\gamma>0$ is the damping coefficient, $\omega_0$ stands for the undamped oscillator angular frequency (sometimes called the natural frequency of the oscillator) and $(q_0,\dot{q}_0)$ encode the initial conditions \cite{grigoriu2021linear, Berkeley}. %The physical units of the displacement $x(t)$ depend on the system being considered. For example, for a mass on a spring (or a pendulum) in viscous fluid, when it is usually called \emph{elongation}, it is measured in $[m]$, while for an RLC circuit it could either be voltage, or current, or charge. In contrast, the units of $\gamma$ and $\omega_0$ are $[s^{-1}]$ for all systems described with the SDOF model.
The form of the solution to the equation \eqref{DHOeq} depends on the relationship between $\gamma$ and $\omega_0$, producing three possible regimes \cite{grigoriu2021linear, Berkeley}: under-damped ($\gamma<\omega_0$), critically damped ($\gamma=\omega_0$) and over-damped ($\gamma>\omega_0$) regime. 

Here we would like to point out that, although it is natural to classify the solution of SDOF into three regimes depending on the value of $\gamma$, we can actually take one form of the solution as a unique solution valid for all values of $\gamma>0$, $\gamma \ne \omega_0$, 
\begin{equation}
q(t)=e^{-\gamma t}\left(q_0\cos(\omega t)+\frac{\dot{q}_0+\gamma q_0}{\omega}\sin(\omega t)\right)\,,
\label{xud}
\end{equation}
where $\omega=\sqrt{\omega_0^2-\gamma^2}$ is the (complex valued) damped angular frequency. In order to describe the critically damped regime, one can take the limit $\gamma\rightarrow\omega_0$ of the solution \eqref{xud} to obtain the general solution of the critically damped regime \cite{Berkeley}. Therefore, in order to calculate the energy and the time integral of the energy, we do not need to perform separate calculations for all three regimes, but a single calculation using the displacement given by \eqref{xud} and the velocity
\begin{equation}
\dot{q}(t)=e^{-\gamma t}\left(\dot{q}_0\cos(\omega t)-\frac{\gamma \dot{q}_0+\omega_0^2 q_0}{\omega}\sin(\omega t)\right)\,.
\label{brzina1D}
\end{equation}
%

%As a typical example of an SDOF system, we will consider a mass $m$ attached to a spring of stiffness $k$ and immersed in a viscous fluid. In this case, $\omega_0=\sqrt{k/m}$. %the energy of the system is given by
%
%\begin{equation}
%E(t)=\frac{m\dot{x}(t)^2}{2}+\frac{m\omega_0^2x(t)^2}{2}\,,
%\label{Et}
%\end{equation}
%
%where $\omega_0=\sqrt{k/m}$.
%For later convenience, we define new \emph{displacement} and \emph{velocity} as 
%
%\begin{equation}
%\begin{split}
%q(t)=\sqrt{\frac{m}{2}}x(t)=e^{-\gamma t}\left(q_0\cos(\omega t)+\frac{\dot{q}_0+\gamma q_0}{\omega}\sin(\omega t)\right)
%\\\dot{q}(t)=\sqrt{\frac{m}{2}}\dot{x}(t)=e^{-\gamma t}\left(\dot{q}_0\cos(\omega t)-\frac{\gamma \dot{q}_0+\omega_0^2 q_0}{\omega}\sin(\omega t)\right)
%\label{cord1D}
%\end{split}
%\end{equation}
%
%respectively, where $q_0=x_0\sqrt{m/2}$ denotes the \emph{initial displacement} and $\dot{q}_0=v_0\sqrt{m/2}$ denotes the \emph{initial velocity}. In terms of \eqref{cord1D}, 
We take that the displacement and velocity are normalised so that the kinetic and potential energy of the system can be written as $E_K(t)=\dot{q}(t)^2$ and $E_P(t)=\omega_0^2q(t)^2$ respectively. Thus, the total energy of the system is 
\begin{equation}
E(t)=E_K(t)+E_P(t)=\dot{q}(t)^2+\omega_0^2q(t)^2\,.
\label{newEt}
\end{equation}
Initially, i.e. at $t=0$, system has energy $E_0=E_{0P}+E_{0K}$, where $E_{0P}=\omega_0^2q_0^2$ and $E_{0K}=\dot{q}_0^2$ are initial potential and initial kinetic energy respectively. All possible initial conditions, with initial energy $E_0$, can be expressed in polar coordinates with constant radius $r=\sqrt{E_0}$ and angle $\theta=\arctan\left(\frac{\dot{q}_0}{\omega_0q_0}\right)$ \cite{Lelas2}, i.e. we can write   
\begin{equation}
\begin{split}
\omega_0q_0=r\cos\theta\\
\dot{q}_0=r\sin\theta\,.
\label{polar}
\end{split}
\end{equation}
Thus, relations \eqref{polar} describe a circle in $(\omega_0q_0,\dot{q}_0)$ coordinate system, and the angle $\theta$ gives us the fraction of potential energy in the initial energy, i.e. $E_{0P}/E_0=\cos^2\theta$, and the fraction of kinetic energy in the initial energy, i.e. $E_{0K}/E_0=\sin^2\theta$.

Using \eqref{xud}, \eqref{brzina1D} and \eqref{polar} in \eqref{newEt}, we obtain the energy   
\begin{equation}
E(\gamma,t,\theta)=E_0e^{-2\gamma t}\left( \cos^2(\omega t)+\gamma\cos2\theta\frac{\sin(2\omega t)}{\omega}+\left(\omega_0^2+\gamma^2+2\omega_0\gamma\sin2\theta\right)\frac{\sin^2(\omega t)}{\omega^2}\right)\,
\label{Et2}
\end{equation}
as a function of $\theta$, instead of $q_0$ and $\dot{q}_0$. Expression \eqref{Et2} is valid for both under-damped and over-damped regime, and the energy of the critically damped regime is obtained easily by taking the $\gamma\rightarrow\omega_0$ limit of the energy \eqref{Et2}. Now we integrate energy \eqref{Et2} over all time, i.e.
\begin{equation}
I(\gamma,\theta)=\int_0^{\infty}E(\gamma,t,\theta )dt\,,
\label{Int1}
\end{equation}
and obtain
\begin{equation}
I(\gamma,\theta)=\frac{E_0}{2\omega_0}\left(\frac{\omega_0}{\gamma}+\frac{2\gamma}{\omega_0}\cos^2\theta+\sin2\theta\right)\,.
\label{Int12}
\end{equation}
Integral \eqref{Int12} is valid for all three regimes, i.e. for any $\gamma>0$.

We want to determine the damping that is optimal according to some criterion for all initial conditions with initial energy $E_0$ or for some given set of initial conditions with initial energy $E_0$. For this purpose, we will analyze the behavior of the energy \eqref{Et2} averaged over a set of initial conditions with $\theta\in[\alpha, \beta]$, i.e. the \emph{average energy} given by
\begin{equation}
\overline{E}(\gamma,t,\alpha, \beta)=E_0e^{-2\gamma t}\left( \cos^2(\omega t)+\gamma\,\overline{\cos2\theta}_{\alpha\beta}\frac{\sin(2\omega t)}{\omega}+\left(\omega_0^2+\gamma^2+2\omega_0\gamma\,\overline{\sin2\theta}_{\alpha\beta}\right)\frac{\sin^2(\omega t)}{\omega^2}\right)\,,
\label{Et2av1}
\end{equation}
and the behavior of the energy integral \eqref{Int12} averaged over a set of initial conditions, i.e. the \emph{average energy integral} given by
\begin{equation}
\overline{I}(\gamma, \alpha, \beta)=\frac{E_0}{2\omega_0}\left(\frac{\omega_0}{\gamma}+\frac{2\gamma}{\omega_0}\,\overline{\cos^2\theta}_{\alpha\beta}+\overline{\sin2\theta}_{\alpha\beta}\right)\,,
\label{Int12av1}
\end{equation}
where we used notation 
\begin{equation}
\overline{f(\theta)}_{\alpha\beta}=\frac{1}{\beta-\alpha}\int_{\alpha}^{\beta}f(\theta)d\theta\,\,\,\,\text{with}\,\,\,\,\lbrace f(\theta)\rbrace=\lbrace\cos2\theta,\sin2\theta,\cos^2\theta\rbrace.
\label{integrals}
\end{equation}
We note here that the energy \eqref{Et2} is periodic in $\theta$ with period $\pi$, i.e. it is invariant to a simultaneous change of the signs of the initial displacement and the initial velocity. Thus, both average energy \eqref{Et2av1} and average energy integral \eqref{Int12av1} are invariant to $[\alpha, \beta]\rightarrow[\alpha+\pi, \beta+\pi]$.

%There are infinitely many different sets of initial conditions corresponding to the state with initial energy $E_0$, i.e. infinitely many different choices for $\alpha$ and $\beta$. The average over all initial conditions is obtained if we take $\alpha=0$ and $\beta=2\pi$, while, e.g., the average over a set of initial conditions with positive displacements only and initial potential energy greater than or equal to the initial kinetic energy (i.e. $q_0>0$ and $E_{0P}\geq E_{0K}$) is obtained if we take $\alpha=-\pi/4$ and $\beta=\pi/4$, etc.
We will consider three sets of initial conditions that correspond to the same initial energy, but differ in the ratio of initial potential and initial kinetic energy of the system:
\begin{itemize}
\item The set of all initial conditions, i.e. $\theta\in[0, 2\pi]$.   
\item The set of initial conditions with $E_{0P}\geq E_{0K}$, i.e. $\theta\in[-\pi/4,\pi/4]\cup\left[3\pi/4,5\pi/4\right]$. Due to the periodicity of \eqref{Et2}, it is sufficient to consider $\theta\in[-\pi/4,\pi/4]$.   
\item The set of initial conditions with $E_{0K}\geq E_{0P}$, i.e. $\theta\in[\pi/4,3\pi/4]\cup\left[5\pi/4,7\pi/4\right]$. It is sufficient to consider $\theta\in[\pi/4,3\pi/4]$.  
\end{itemize}
%
%Thus, throughout this paper we consider sets $\lbrace[\alpha,\beta]\rbrace=\lbrace[0,2\pi],[-\pi/4,\pi/4],[\pi/4,3\pi/4]\rbrace$.

In this paper, we study optimal damping with respect to sets of initial conditions with the same initial energy, and, in general, initial conditions may differ in initial energy as well. This can be taken into account by making an additional integration (averaging) over some range of initial energies of interest in relations \eqref{Et2av1} and \eqref{Int12av1}. Furthermore, in practice some initial conditions may have a higher frequency of occurrence than others. This can be taken into account by introducing weight functions in relations \eqref{Et2av1} and \eqref{Int12av1}. In this paper, we take that all initial conditions have the same weight. 

\subsection{Minimization of the average energy integral} 
\label{1Daverage}

If we consider that the optimal damping, adapted to the set of initial conditions with $\theta\in[\alpha, \beta]$, is the one for which the average energy integral \eqref{Int12av1} is minimal, we can easily determine the corresponding optimal damping coefficient $\gamma_{\opt}$ from the condition 
\begin{equation}
\frac{\partial \overline{I}(\gamma, \alpha, \beta)}{\partial\gamma}\bigg|_{\gamma_{\opt}}=0\,,
\label{1min}
\end{equation}
and we obtain
\begin{equation}
\gamma_{\opt}=\sqrt{\frac{1}{2\overline{\cos^2\theta}_{\alpha\beta}}}\omega_0=\sqrt{\frac{2(\beta-\alpha)}{2(\beta-\alpha)+\sin2\beta-\sin2\alpha}}\omega_0\,.
\label{gopt}
\end{equation}
For easier comparison, we show the results for optimal damping coefficients \eqref{gopt} in Fig.\ \ref{fig:1D}, where we also show the behaviour of the optimal damping coefficients obtained with the new method, i.e. by considering the fastest drop of the average energy.

Before closing this subsection, we point out that the relation \eqref{gopt} was recently derived in \cite{Lelas2} and the results for the optimal damping coefficients of the SDOF system, given by the minimisation of the average energy integral for different sets of initial conditions were already analyzed there. We repeated them here for completeness, while we extend the analysis to MDOF systems in section \ref{MDOF}. 

\subsection{Fastest drop of the average energy} 
\label{1DaverageE}

%From a theoretical perspective, systems with viscous damping asymptotically approach the equilibrium state and never reach it exactly. In nature and in experiments, these systems reach the equilibrium state which is not an exact zero energy state, but rather a state in which the energy of the system has decreased to the level of the energy imparted to the system by the surrounding noise, or to the energy resolution of the measuring apparatus. Furthermore, in the context of vibration control, it is important to determine the damping for which the vibration energy drops the fastest to a level that no longer poses a problem.
We define that the vibrations, which started with some specific initial condition $\theta$, have decreased to a negligible, or acceptable, level for times $t>\tau$ such that  
\begin{equation}
\frac{E(\gamma,\tau,\theta)}{E_0}=10^{-\delta}\,,
\label{conditiontau}
\end{equation}
where $E(\gamma,\tau,\theta)$ is the energy of the system at $t=\tau$, and $\delta>0$ is a dimensionless parameter that defines what fraction of the initial energy is left in the system. This reasoning has recently been used to determine the optimal damping of the SDOF system \cite{Lelas}, and the MDOF systems with MPD \cite{Lelas2}, as the damping for which the energy drops the fastest to a given threshold value. In order to find optimal damping adapted to a set of initial conditions $\theta\in[\alpha,\beta]$, we consider as optimal the damping for which the average energy $\overline{E}(\gamma, t, \alpha, \beta)$, given by \eqref{Et2av1}, drops to a given threshold $E_{th}=10^{-\delta}E_0$ the fastest and we denote it with $\tilde{\gamma}_{\opt}$. 

\begin{figure}[h!t!]
\begin{center}
\includegraphics[width=0.48\textwidth]{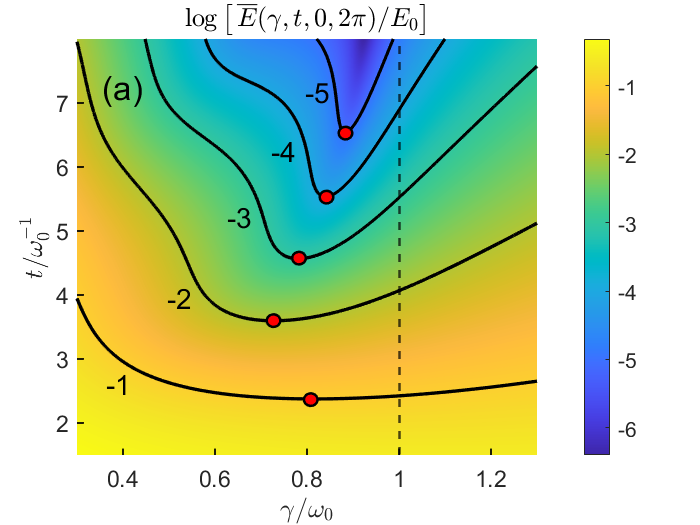}
\includegraphics[width=0.48\textwidth]{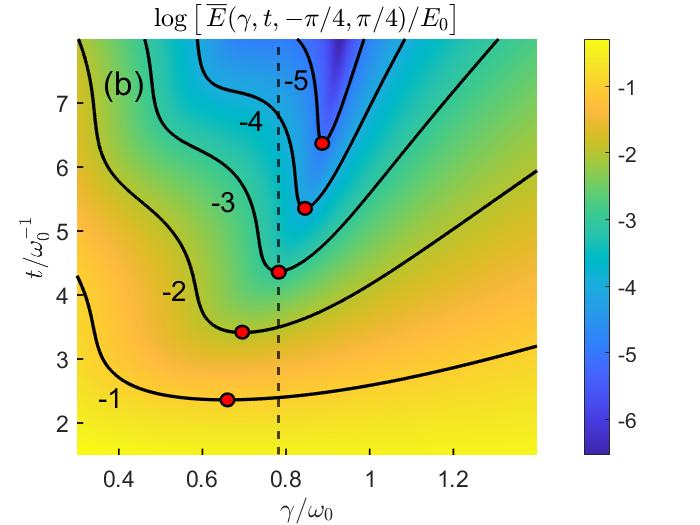}
\includegraphics[width=0.48\textwidth]{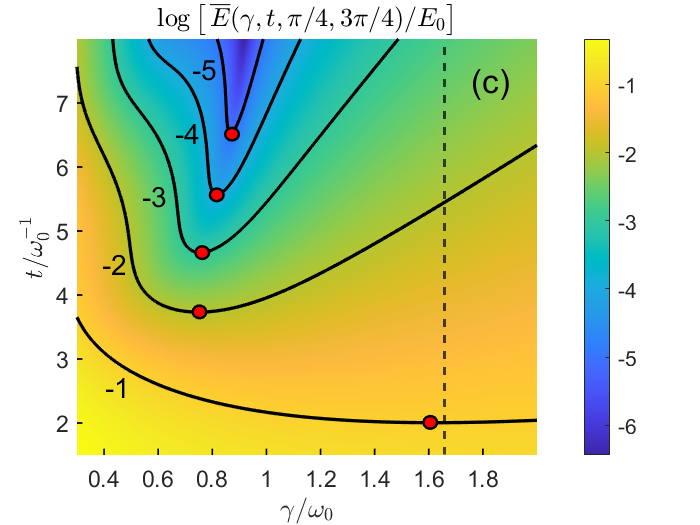}
\end{center}
\caption{The base $10$ logarithm of the average energy \eqref{Et2av1} to initial energy ratio of the SDOF system for (a) $\theta\in[0,2\pi]$, (b) $\theta\in[-\pi/4,\pi/4]$ and (c) $\theta\in[\pi/4,3\pi/4]$. Contour lines denote points with $\overline{E}(\gamma,t,\alpha,\beta)/E_{0}=\lbrace 10^{-1}, 10^{-2},10^{-3},10^{-4},10^{-5}\rbrace$ respectively, as indicated by the numbers placed to the left of each contour line. Filled red circles denote points with $\gamma=\tilde{\gamma}_{\opt}$. The black dashed vertical lines are positioned at $\gamma_{\opt}$ given by \eqref{gopt} for the corresponding initial conditions.} %(a) $\gamma_{\opt}=\omega_{0}$, (b) $\gamma_{\opt}=0.78\omega_{0}$ and (c) $\gamma_{\opt}=1.66\omega_{0}$, given by \eqref{gopt}.}
\label{fig:1D}
\end{figure}

In Fig.\ \ref{fig:1D}(a) we show the base $10$ logarithm of the average energy \eqref{Et2av1} to initial energy ratio for $\theta\in[0,2\pi]$, i.e. $\log[\overline{E}(\gamma,t,0,2\pi)/E_0]$, as a function of $\gamma$ and $t$. Five black contour lines denote points with $\overline{E}(\gamma,t,0,2\pi)/E_0=\lbrace10^{-1},10^{-2},10^{-3},10^{-4},10^{-5}\rbrace$ respectively, as indicated by the numbers placed to the left of each contour line. Each contour line has a unique point closest to the $\gamma$ axis, i.e. corresponding to the damping coefficient $\tilde{\gamma}_{\opt}$ for which that energy level is reached the fastest. In Fig.\ \ref{fig:1D}(b) and (c), we show $\log[\overline{E}(\gamma,t,-\pi/4,\pi/4)/E_0]$ and $\log[\overline{E}(\gamma,t,\pi/4,3\pi/4)/E_0]$ respectively, and we see qualitatively the same behavior as in Fig.\ \ref{fig:1D}(a). Therefore, for all three sets of initial conditions, we can easily read out the optimal damping coefficients $\tilde{\gamma}_{\opt}$. Rather than overwhelming the reader with a bunch of numerical results for all the different thresholds $E_{th}$ and for all three sets of initial conditions, on each contour line we have marked with filled red circle the point closest to the $\gamma$ axis. The black dashed vertical lines are positioned at the optimal damping coefficients $\gamma_{\opt}$ given by \eqref{gopt}. This way we can visually analyze the relationship between $\tilde{\gamma}_{\opt}$ and $\gamma_{\opt}$. 
%The results are given in Table \ref{table1}-\ref{table3} (rounded to two decimal places).
%

It can be inferred from the results presented in Fig.\ \ref{fig:1D} that for all three sets of initial conditions the optimal damping coefficients $\tilde{\gamma}_{\opt}$ converge to critical damping if lower and lower energy thresholds are considered, i.e. $\tilde{\gamma}_{\opt}\rightarrow\omega_0$ for $E_{th}\rightarrow 0$. On the other hand, we see that the optimal damping coefficients $\gamma_{\opt}$ are significantly different depending on the considered set of initial conditions. We can see that $\tilde{\gamma}_{\opt}$ converges to $\gamma_{\opt}$, for $E_{th}\rightarrow 0$, only for the set of all initial conditions. 

%For the set with $E_{0P}\geq E_{0K}$, the two methods agree excellent, i.e. $\tilde{\gamma}_{\opt}=\gamma_{\opt}$, for the threshold $E_{th}=10^{-3}E_{0}$, and as we go towards smaller/bigger threshold values, $\tilde{\gamma}_{\opt}$ becomes larger/smaller than $\gamma_{\opt}$. For the set with $E_{0K}\geq E_{0P}$, the two methods give similar results for the threshold $E_{th}=10^{-1}E_{0}$, while $\tilde{\gamma}_{\opt}$ is significantly smaller than $\gamma_{\opt}$ for smaller threshold values.

%For all three sets of initial conditions, we can see that the average energy drops approximately at the same time to the threshold $E_{th}=10^{-1}E_{0}$, for a wider range of damping coefficients. Therefore, if we are considering $E_{th}=10^{-1}E_{0}$, even though there are differences in magnitude between $\tilde{\gamma}_{\opt}$ and $\gamma_{\opt}$, the differences in the average energy dissipation rate for these two coefficients are insignificant. %For lower thresholds, the agreement between $\tilde{\gamma}_{\opt}$ and 2 is best for initial condition set 1, while the differences between 1 and 2 are most pronounced for initial condition set 3.
%.....nastavit ce se, ovdje cu jos malo opisat ponasanje koje vidimo na slikama.... 

%\section{2-DOF systems}
\section{MDOF systems with MPD: minimization of the average energy integral and fastest drop of the average energy}
\label{MDOF}

\subsection{Short review of the model, systematization of initial conditions, average energy and average energy integral}
\label{2Dsubsec1}

\begin{figure}[h!t!]
\begin{center}
\includegraphics[width=1\textwidth]{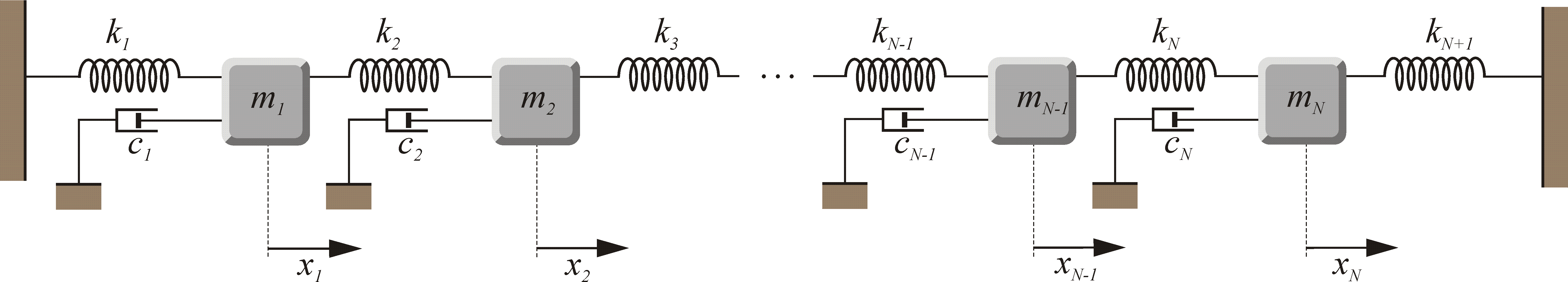}
\end{center}
\caption{Schematic figure of a MDOF system with $N$ degrees of freedom.}
\label{fig:skicaND}
\end{figure}

%The corresponding equation of motion are
%
%\begin{equation}
%\begin{split}
%m_i\ddot x_i(t)=-c_i\dot x_i(t)-k_ix_i(t)-k_{i+1}\left(x_i(t)-x_{i+1}(t)\right)\,,
%\\m_2\ddot x_2(t)=-c_2\dot x_2(t)-k_3x_2(t)+k_2\left(x_1(t)-x_2(t)\right)\,.
%\label{eq2D}
%\end{split}
%\end{equation}
%
%In order to be able to use modal analysis
%We will consider MPD \cite{muravyov1998geometrical}, i.e. masses $\lbrace m_1,m_2\rbrace$, spring constants $\lbrace k_1,k_2,k_3\rbrace$, and dampers $\lbrace c_1,c_2\rbrace$ can in general be mutually different but the condition $c_1/m_1=c_2/m_2$ holds.

Here we consider MDOF system with $N$ degrees of freedom, shown schematically in Fig.\ \ref{fig:skicaND}, with MPD \cite{muravyov1998geometrical}, i.e. masses $\lbrace m_1,m_2,...,m_N\rbrace$, spring constants $\lbrace k_1,k_2,...,k_{N+1}\rbrace$, and dampers $\lbrace c_1,c_2,...,c_N\rbrace$ can in general be mutually different but the condition $c_i/m_i=2\gamma$ holds for any $i=\lbrace1,...,N\rbrace$, where $\gamma$ is the damping coefficient. In this case we can use modal analysis \cite{fu2001modal, grigoriu2021linear} and the equation of motion for the $i$-th mode can be written as
\begin{equation}
%\begin{split}
\ddot q_i(t)+2\gamma\dot q_i(t)+\omega_{0i}^2q_i(t)=0\,
%\\\ddot q_2(t)+2\gamma\dot q_2(t)+\omega_{02}^2q_2(t)=0\,,
\label{eq2Dnorm}
%\end{split}
\end{equation}
where $q_i(t)$ and $\omega_{0i}$, for $i=\lbrace 1,2,...,N\rbrace$, denote the modal coordinates and undamped modal angular frequencies of the $i$-th mode \cite{grigoriu2021linear}. In the analysis that we will carry out in this section, we will not need the explicit connection of modal coordinates $q_i(t)$ and mass coordinates, i.e. mass displacements $x_i(t)$. Similarly as in Section \ref{1Dsection} (see \eqref{xud}), the general solution for the $i$-th mode can be written as
\begin{equation}
q_i(t)=e^{-\gamma t}\left(q_{0i}\cos(\omega_i t)+\frac{\dot q_{0i}+\gamma q_{0i}}{\omega_i}\sin(\omega_i t)\right)\,,%\,\textrm{with}\,\,i=\lbrace1,2\rbrace\,,
\label{mod2D}
\end{equation}
where $\omega_i=\sqrt{\omega_{0i}^2-\gamma^2}$ is the damped modal angular frequency, and $q_i(0)=q_{0i}$ and $\dot q_i(0)=\dot q_{0i}$ are the initial conditions of the $i$-th mode. Thus, the reasoning and the results presented in Section \ref{1Dsection}, with some adjustments, can by applied for the analysis of the MDOF system we are considering here.

We take that the modal coordinates are normalised so that the total energy of the system can be written as
\begin{equation}
E(t)=\sum_{i=1}^NE_i(t)=\sum_{i=1}^N\left(\dot q_i(t)^2+\omega_{0i}^2q_i(t)^2\right)\,
\label{energyND}
\end{equation}
where $E_i(t)$ in \eqref{energyND} denotes the energy of the $i$-th mode. Total energy at $t=0$ is given by 
\begin{equation}
E_0=\sum_{i=1}^NE_{0i}=\sum_{i=1}^N\left(E_{0Ki}+E_{0Pi}\right)=\sum_{i=1}^N\left(\dot q_{0i}^2+\omega_{0i}^2q_{0i}^2\right)\,,
\label{energy0ND}
\end{equation}
where $E_{0i}$ denotes the initial energy of the $i$-th mode, $E_{0Ki}=\dot q_{0i}^2$ and $E_{0Pi}=\omega_{0i}^2q_{0i}^2$ denote initial kinetic and initial potential energy of the $i$-th mode.

%All possible initial conditions with the same initial energy \eqref{energy0ND} can be expressed as in the SDOF case, but with $N$ pairs of polar coordinates, one pair for each mode. For the $i$-th mode we have radius $r_i=\sqrt{E_{0i}}$ and angle $\theta_i=\arctan\left(\frac{\dot q_{0i}}{\omega_{0i}q_{0i}}\right)$ \cite{Lelas2}. Thus, by using relation \eqref{Et2} for each of the modes, we can write the total energy of the system, i.e. \eqref{energyND}, as
All possible initial conditions with the same initial energy \eqref{energy0ND} can be expressed as in the SDOF case, but with $N$ pairs of polar coordinates, one pair for each mode, i.e. we can write 
\begin{equation}
\begin{split}
\omega_{0i}q_{0i}=r_i\cos\theta_i\\
\dot{q}_{0i}=r_i\sin\theta_i\,
\label{polarN}
\end{split}
\end{equation}
for the initial conditions of the $i$-th mode, where we have radius $r_i=\sqrt{E_{0i}}$ and angle $\theta_i=\arctan\left(\frac{\dot q_{0i}}{\omega_{0i}q_{0i}}\right)$ \cite{Lelas2}. Thus, by using relation \eqref{Et2} for each of the modes, i.e. for each $E_i(t)$ in \eqref{energyND}, we can write the total energy of the system as
\begin{equation}
E(\gamma,t,\lbrace E_{0i}\rbrace, \lbrace\theta_i\rbrace)=e^{-2\gamma t}\sum_{i=1}^NE_{0i}\left( \cos^2(\omega_i t)+\gamma\cos2\theta_i\frac{\sin(2\omega_i t)}{\omega_i}+\left(\omega_{0i}^2+\gamma^2+2\omega_{0i}\gamma\sin2\theta_i\right)\frac{\sin^2(\omega_i t)}{\omega_i^2}\right)\,.
\label{EipolarTOT}
\end{equation}

We consider here initial conditions with all possible distributions of total initial energy over initial energies of the modes, i.e. $E_{0i}\in[0,E_0]$ $\forall i$ and \eqref{energy0ND} hold, and with distributions of initial potential and initial kinetic energy within the modes determined by angles $\theta_i\in[\alpha, \beta]$ $\forall i$. In this case, the average value of initial energy of the $i$-th mode is $\overline{E}_{0i}=E_0/N$ (see relation (61) and Appendix A in \cite{Lelas2}), and averaging energy \eqref{EipolarTOT} over polar angles involves integrals of the form
\begin{equation}
\frac{1}{(\beta-\alpha)^N}\int_{\alpha}^{\beta}d\theta_1...\int_{\alpha}^{\beta}d\theta_i...\int_{\alpha}^{\beta}d\theta_N\,f(\theta_i)=\frac{1}{(\beta-\alpha)}\int_{\alpha}^{\beta}d\theta_i\,f(\theta_i)\,.
\label{integralsMDOF}
\end{equation}
Thus, we get 
\begin{equation}
\overline{E}(\gamma,t,\alpha,\beta)=\frac{E_0}{N}e^{-2\gamma t}\sum_{i=1}^N\left( \cos^2(\omega_i t)+\gamma\,\overline{\cos2\theta}_{\alpha\beta}\frac{\sin(2\omega_i t)}{\omega_i}+\left(\omega_{0i}^2+\gamma^2+2\omega_{0i}\gamma\,\overline{\sin2\theta}_{\alpha\beta}\right)\frac{\sin^2(\omega_i t)}{\omega_i^2}\right)\,
\label{EipolarTOTAV}
\end{equation} 
for the average energy, and 
\begin{equation}
\overline{I}(\gamma, \alpha, \beta)=\frac{E_0}{N}\sum_{i=1}^N\frac{1}{2\omega_{0i}}\left(\frac{\omega_{0i}}{\gamma}+\frac{2\gamma}{\omega_{0i}}\,\overline{\cos^2\theta}_{\alpha\beta}+\overline{\sin2\theta}_{\alpha\beta}\right)\,
\label{Int12av2D}
\end{equation}
for the average energy integral. In \eqref{EipolarTOTAV} and \eqref{Int12av2D}, we use the same notation for average values of trigonometric functions as in \eqref{integrals}. 

We note here that, at least for MDOF systems with MPD, averaging over the set of initial conditions with all polar angles equal, i.e. with $\theta_i=\theta$ $\forall i$ and $\theta\in[\alpha, \beta]$, gives also \eqref{EipolarTOTAV} for the average energy and \eqref{Int12av2D} for the average energy integral (see \eqref{integralsMDOF}).

\subsection{Minimization of the average energy integral}

\begin{figure}[h!t!]
\begin{center}
\includegraphics[width=0.48\textwidth]{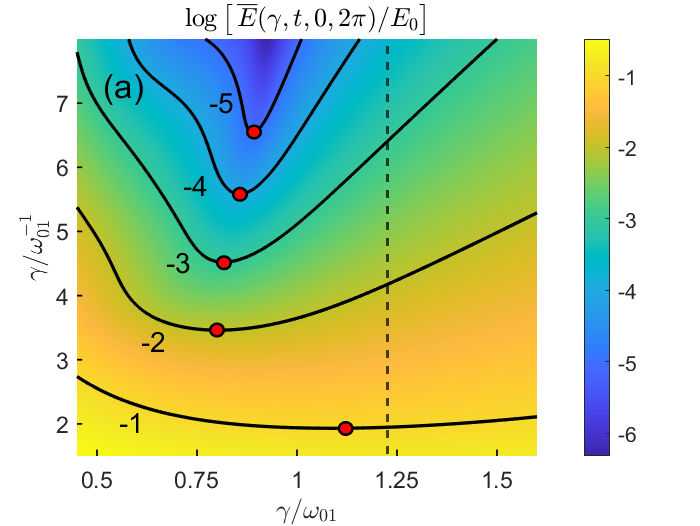}
\includegraphics[width=0.48\textwidth]{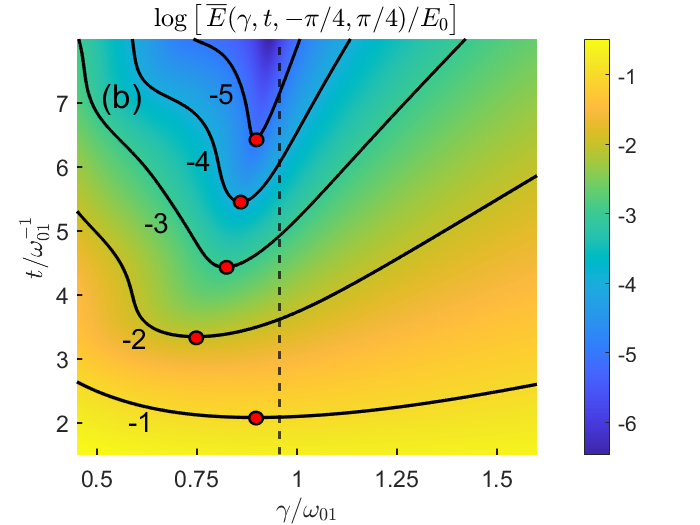}
\includegraphics[width=0.48\textwidth]{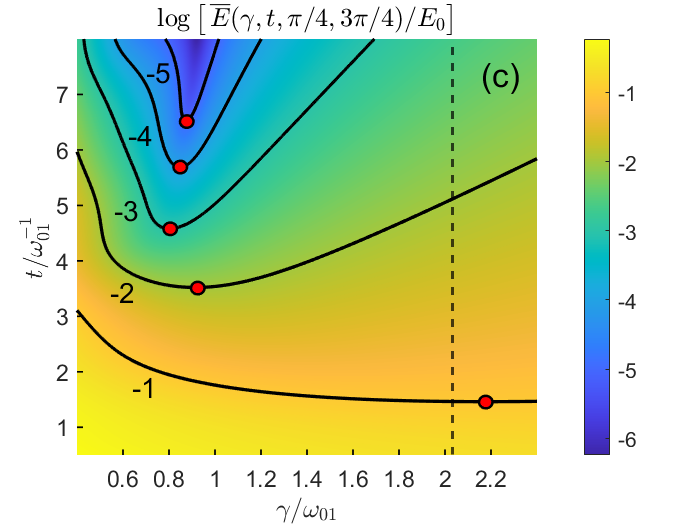}
\end{center}
\caption{The base $10$ logarithm of the average energy \eqref{EipolarTOTAV} to initial energy ratio of the 2-DOF system for (a) $\theta_i\in[0,2\pi]$, (b) $\theta_i\in[-\pi/4,\pi/4]$ and (c) $\theta_i\in[\pi/4,3\pi/4]$ for $i=1,2$. Contour lines denote points with $\overline{E}(\gamma,t,\alpha,\beta)/E_{0}=\lbrace 10^{-1}, 10^{-2},10^{-3},10^{-4},10^{-5}\rbrace$ respectively, as indicated by the numbers placed to the left of each contour line. Filled red circles denote points with $\gamma=\tilde{\gamma}_{\opt}$. The black dashed vertical lines are positioned at the optimal damping coefficients $\gamma_{\opt}$ given by \eqref{gammaopt} for the 2-DOF system for the corresponding initial conditions.} % (a) $\gamma_{\opt}=1.23\omega_{01}$, (b) $\gamma_{\opt}=0.96\omega_{01}$ and (c) $\gamma_{\opt}=2.03\omega_{01}$, given by \eqref{gammaopt} for the 2-DOF system.}
\label{fig:2Dsvi}
\end{figure}

\begin{figure}[h!t!]
\begin{center}
\includegraphics[width=0.48\textwidth]{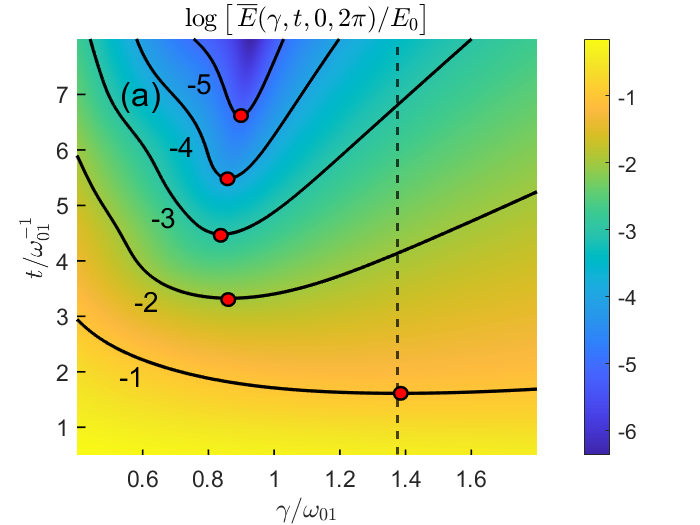}
\includegraphics[width=0.48\textwidth]{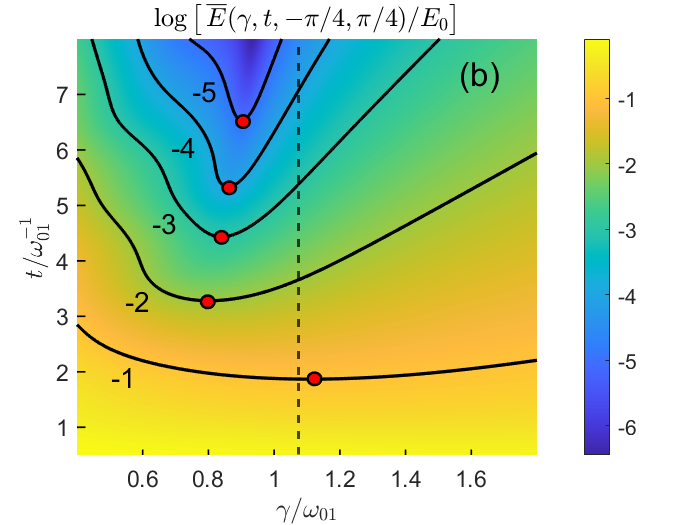}
\includegraphics[width=0.48\textwidth]{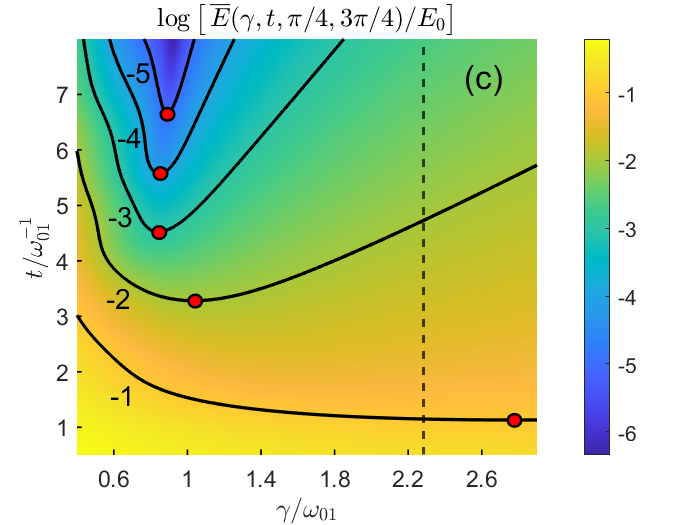}
\end{center}
\caption{The base $10$ logarithm of the average energy \eqref{EipolarTOTAV} to initial energy ratio of the 3-DOF system for (a) $\theta_i\in[0,2\pi]$, (b) $\theta_i\in[-\pi/4,\pi/4]$ and (c) $\theta_i\in[\pi/4,3\pi/4]$ for $i=1,2,3$. Contour lines denote points with $\overline{E}(\gamma,t,\alpha,\beta)/E_{0}=\lbrace 10^{-1}, 10^{-2},10^{-3},10^{-4},10^{-5}\rbrace$ respectively, as indicated by the numbers placed to the left of each contour line. Filled red circles denote points with $\gamma=\tilde{\gamma}_{\opt}$. The black dashed vertical lines are positioned at the optimal damping coefficients $\gamma_{\opt}$ given by \eqref{gammaopt} for the 3-DOF system for the corresponding initial conditions.}% The black dashed vertical lines are positioned at the optimal damping coefficients (a) $\gamma_{\opt}=1.38\omega_{01}$, (b) $\gamma_{\opt}=1.08\omega_{01}$ and (c) $\gamma_{\opt}=2.28\omega_{01}$, given by \eqref{gammaopt} for the 3-DOF system.}
\label{fig:3Dsvi}
\end{figure}

We differentiate relation \eqref{Int12av2D} with respect to $\gamma$, equate it to zero and get
\begin{equation}
\gamma_{\opt}=\left(\frac{N}{2\,\overline{\cos^2\theta}_{\alpha\beta}}\right)^{1/2}\left(\sum_{i=1}^N\frac{1}{\omega_{0i}^2}\right)^{-1/2}\,.
\label{gammaopt}
\end{equation}
As quantitative examples, we consider the results given by \eqref{gammaopt} for 2-DOF and 3-DOF systems with masses $m$, spring constants $k$, and dampers $c$. In this case, the corresponding natural (undamped) angular frequencies of the modes are \cite{grigoriu2021linear}
\begin{equation}
\omega_{0i}=2\omega_0\sin\left(\frac{i\pi}{2(N+1)}\right)\,,\,\,\text{with}\,\,i=\lbrace1,...,N\rbrace\,,
\label{omegeN}
\end{equation}
where $\omega_0=\sqrt{k/m}$, and $N$ is the number of degrees of freedom. For easier comparison of the two methods, we show the results for $\gamma_{\opt}$ in Fig.\ \ref{fig:2Dsvi} and Fig.\ \ref{fig:3Dsvi}, together with the results for $\tilde{\gamma}_{\opt}$. 

\subsection{Fastest drop of the average energy} 
\label{2DaverageE}

%\subsubsection{2-DOF system}
In Fig.\ \ref{fig:2Dsvi} and Fig.\ \ref{fig:3Dsvi} we show the base $10$ logarithm of the average energy \eqref{EipolarTOTAV} to initial energy ratio of the 2-DOF and 3-DOF systems respectively, for the three chosen sets of initial conditions. Similarly as in Fig.\ \ref{fig:1D}, on all figures contour lines denote points with $\overline{E}(\gamma,t,\alpha,\beta)/E_{0}=\lbrace 10^{-1}, 10^{-2},10^{-3},10^{-4},10^{-5}\rbrace$ respectively, as indicated by the numbers placed to the left of each contour line. The filled red circles denote the points on the contour lines that are closest to the $\gamma$ axis, i.e. corresponding to the optimal damping coefficient $\tilde{\gamma}_{\opt}$. On all figures, the black dashed vertical lines are positioned at the optimal damping coefficients $\gamma_{\opt}$ given by \eqref{gammaopt}. The damping coefficients of the 2-DOF and 3-DOF systems are given in the corresponding $\omega_{01}$ units, and time in $\omega_{01}^{-1}$ units. 

We can see in Fig.\ \ref{fig:2Dsvi} and Fig.\ \ref{fig:3Dsvi} that the optimal damping coefficients $\tilde{\gamma}_{\opt}$ converge to critical damping of the first mode, i.e. to undamped angular frequency $\omega_{01}$, if $E_{th}\rightarrow0$, for all three sets of initial conditions. In contrast to the SDOF case, we see that in the 2-DOF and 3-DOF case the optimal damping coefficients $\tilde{\gamma}_{\opt}$ do not converge to $\gamma_{\opt}$ for the set of all initial conditions, if $E_{th}\rightarrow0$, and we see that the differences between $\tilde{\gamma}_{\opt}$ and $\gamma_{\opt}$ become more pronounced as we go from the 2-DOF case to the 3-DOF case.

As the number of degrees of freedom $N$ further increases, the optimal damping coefficient $\tilde{\gamma}_{\opt}$ would still converge to $\omega_{01}$, if we would consider $E_{th}\rightarrow0$. The reason for this is that the behavior of the average energy \eqref{EipolarTOTAV} becomes dominantly determined by the first mode as the time increases, since all higher modes lose energy faster than the first mode, and the optimal damping of the first mode alone converges to its critical damping if we consider $E_{th}\rightarrow0$. %\cite{Lelas}.%ovaj paragraf se moze analiticki potkrijepiti, kao sto sam radio u AJP clanku, samo primjenjeno na izraz za prosječnu MDOF energiju, pa, ako budu trazili, ubacim.
On the other hand, it was recently shown that the ratio of optimal damping coefficient \eqref{gammaopt} for the set of all initial conditions with the same initial energy and the undamped angular frequency of the first mode, i.e. $\gamma_{\opt}/\omega_{01}$ for $\theta\in[0, 2\pi]$, increases to infinity for $N\rightarrow\infty$ \cite{Lelas2}. From relation \eqref{gammaopt} we see that the same is true for any other set of initial conditions, since the relation \eqref{gammaopt} for the set $\theta\in[\alpha, \beta]$ and the set $\theta\in[0, 2\pi]$ differs only up to the factor. Therefore, if we consider $E_{th}\rightarrow0$, the ratio of the optimal damping coefficients given by the two methods, i.e. $\gamma_{\opt}/\tilde{\gamma}_{\opt}$, increases to infinity as the number of degrees of freedom increases to infinity. We can conclude that shifting the focus from the average energy integral to the average energy leads to significantly different results for the optimal damping adapted to a set of initial conditions.

%\begin{comment}
\section{Minimum of the average settling time} 
\label{averageT}

\subsection{Average settling time of the SDOF system}

In accordance with condition \eqref{conditiontau}, we define the \emph{settling time} of the SDOF system, denoted as $\tau(\gamma,\theta)$, as the time required for the energy of the system \eqref{Et2} to drop to a given threshold value, i.e. $E(\gamma,t,\theta)=E_{th}$ for $t=\tau(\gamma,\theta)$. For a set of initial conditions with $\theta\in[\alpha, \beta]$ we define the \emph{average settling time} as
\begin{equation}
\overline{\tau}(\gamma)=\frac{1}{\beta-\alpha}\int_{\alpha}^{\beta}\tau(\gamma,\theta)d\theta\,.
\label{tauav}
\end{equation}
We will consider here the optimal damping to be the one that minimizes the average settling time and denote it with $\gamma_{\opt}^{\star}$, i.e. $\overline{\tau}(\gamma)\rightarrow\text{min}$ for $\gamma\rightarrow\gamma_{\opt}^{\star}$.

\begin{figure}[h!t!]
\begin{center}
\includegraphics[width=0.45\textwidth]{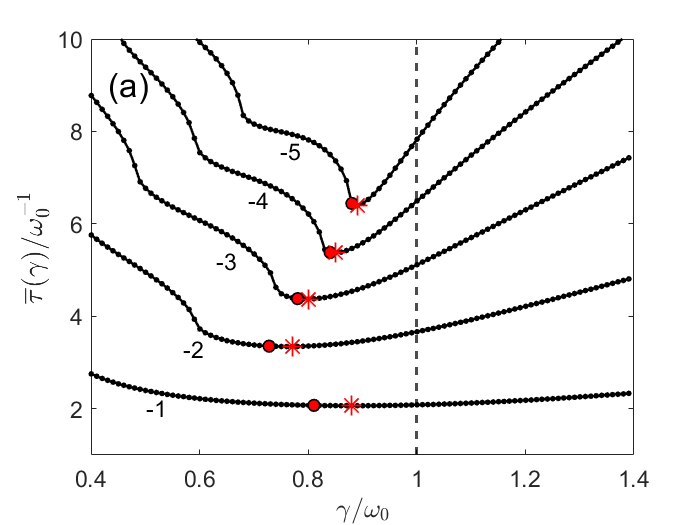}
\includegraphics[width=0.45\textwidth]{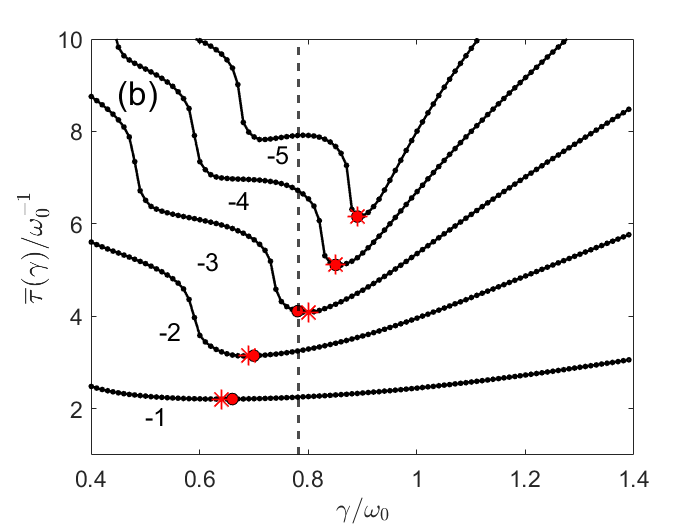}
\includegraphics[width=0.45\textwidth]{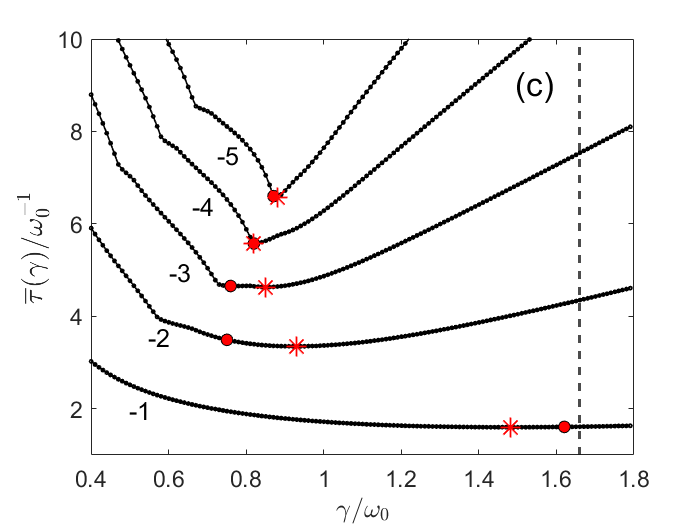}
\end{center}
\caption{Average settling time of the SDOF system, i.e. \eqref{tauav}, for (a) $\theta\in A$, (b) $\theta\in B$ and (c) $\theta\in C$. The numbers $-\delta$, placed next to each curve, indicate the considered threshold $E_{th}=10^{-\delta}E_0$. The red asterisks denote points $\left(\gamma_{\opt}^{\star},\overline{\tau}(\gamma_{\opt}^{\star})\right)$, i.e. points with minimum average settling time. Filled red circles denote points $\left(\tilde{\gamma}_{\opt},\overline{\tau}(\tilde{\gamma}_{\opt})\right)$. The black dashed vertical lines are positioned at $\gamma_{\opt}$.}% The step between the displayed points along the $\gamma$-axis is $\Delta\gamma=0.01\omega_0$.}
\label{fig:1Dsettime}
\end{figure}

Since $\tau(\gamma,\theta)$ cannot be determined analytically, we now replace continuous sets of initial conditions with discrete sets, i.e.  
%
%\begin{itemize}
%\item $\theta\in[0,2\pi]\rightarrow \theta\in A=\lbrace \frac{k\pi}{360}\,|\,k\in\mathbb{Z},0\leq k \leq 719\rbrace$   
%\item $\theta\in[-\pi/4,\pi/4]\rightarrow \theta\in B=\lbrace \frac{k\pi}{360}\,|\,k\in\mathbb{Z},-90\leq k \leq 90\rbrace$.   
%\item $\theta\in[\pi/4,3\pi/4]\rightarrow \theta\in C=\lbrace \frac{k\pi}{360}\,|\,k\in\mathbb{Z},90\leq k \leq 270\rbrace$.  
%\end{itemize}
%
%
\begin{equation}
\begin{split}
\theta\in[0,2\pi]\rightarrow \theta\in A=\lbrace \frac{k\pi}{360}\,|\,k\in\mathbb{Z},0\leq k \leq 359\rbrace\\   
\theta\in[-\pi/4,\pi/4]\rightarrow \theta\in B=\lbrace \frac{k\pi}{360}\,|\,k\in\mathbb{Z},-90\leq k \leq 90\rbrace\\   
\theta\in[\pi/4,3\pi/4]\rightarrow \theta\in C=\lbrace \frac{k\pi}{360}\,|\,k\in\mathbb{Z},90\leq k \leq 270\rbrace\,,
\end{split}
\label{sets}
\end{equation}
where we took into account the periodicity of energy and settling time in $\theta$ in the first line of \eqref{sets}. Thus, using condition $E(\gamma,\tau,\theta)=E_{th}$ we find numerically $\tau(\gamma,\theta)$ for each $\theta$ in the discrete sets \eqref{sets} and for a range of damping coefficients $\gamma$, and calculate the average settling time \eqref{tauav} numerically with $d\theta$ replaced by $\Delta\theta=\pi/360$. The results are shown in Fig.\ \ref{fig:1Dsettime}, where we point out the coordinates of the minimal average settling time (red asterisks), and the values of the average settling time obtained for optimal damping coefficients given by the other two methods, i.e. for $\tilde{\gamma}_{\opt}$ (filled red circle) and $\gamma_{\opt}$ (dashed vertical line), for comparison.

At first glance, it may seem that finding the minimum of the average settling time is a more complicated version of the method presented in subsection \ref{1Daverage} where we considered the fastest drop of the average energy. In order to explain the differences and similarities between these two methods, i.e. between $\gamma_{\opt}^{\star}$ and $\tilde{\gamma}_{\opt}$, and thus at the same time to clarify the results shown in Fig.\ \ref{fig:1Dsettime}, we will now consider two thought experiments.

First thought experiment. Imagine that we have a SDOF system whose dynamics is described well by equation \eqref{DHOeq} and, for simplicity, that we are interested in free vibrations of that system for only two different initial conditions, both of which have the same initial energy $E_0$. We will refer to them as initial condition $a$ and initial condition $b$. First we excite free vibrations with initial condition $a$ and measure the settling time $\tau(\gamma,a)$ needed for the system energy $E(\gamma,t,a)$ to drop to some energy threshold $E_{th}$, which can represent, e.g., acceptable level of vibrational energy, i.e. vibrations with $E(\gamma,t,a)\leq E_{th}$ pose no problem. Afterwards, we either stop the residual vibrations ourselves or simply wait for the system to fully return to the equilibrium state and then excite free vibrations with the initial condition $b$ and measure the settling time $\tau(\gamma,b)$ needed for the system energy $E(\gamma,t,b)$ to drop to the same energy threshold $E_{th}$. The total time the system vibrated with an energy greater than $E_{th}$ during such an event-by-event experiment is $\tau_{tot}(\gamma)=\tau(\gamma,a)+\tau(\gamma,b)$, and the average settling time is $\overline{\tau}(\gamma)=\tau_{tot}(\gamma)/2$. Thus, the optimal damping coefficient  $\gamma_{\opt}^{\star}$ is the one that minimizes  $\overline{\tau}(\gamma)$ and $\tau_{tot}(\gamma)$, i.e. $\overline{\tau}(\gamma)\rightarrow\text{min}$ and $\tau_{tot}(\gamma)\rightarrow\text{min}$ for $\gamma\rightarrow\gamma_{\opt}^{\star}$. 

Second thought experiment. Imagine that we have two independent SDOF systems, both identical to the system in the first thought experiment. We will refer to them as system one and system two, and to both of them together as the total system. We excite at the same time system one with initial condition $a$ and system two with initial condition $b$ and measure the time $T(\gamma)$ needed for the energy of total system $E_{tot}(\gamma,t)=E(\gamma,t,a)+E(\gamma,t,b)$ to drop to $2E_{th}$. We refer to $T(\gamma)$ as the settling time of the total system. The energy of SDOF system averaged over initial conditions $a$ and $b$ is $\overline{E}(\gamma,t)=E_{tot}(\gamma,t)/2$. Average energy $\overline{E}(\gamma,t)$ drops to the level $E_{th}$ at $t=T(\gamma)$, i.e. for the same time interval for which the energy of total system falls from initial energy $2E_0$ to energy $2E_{th}$. Thus, the optimal damping coefficient  $\tilde{\gamma}_{\opt}$ is the one that gives the fastest drop of average energy $\overline{E}(\gamma,t)$ and that minimizes settling time of the total system $T(\gamma)$ in this type of experiment.

It is clear from the two previous paragraphs that $\gamma_{\opt}^{\star}$ and $\tilde{\gamma}_{\opt}$ correspond to different situations, i.e. these two damping coefficients are generally different. Therefore, the quantitative differences between $\gamma_{\opt}^{\star}$ and $\tilde{\gamma}_{\opt}$, which we see in Fig.\ \ref{fig:1Dsettime}, are not the consequence of numerical calculation, i.e. of finite number of initial conditions in discrete sets \eqref{sets}, but are a consequence of the qualitative difference between these two methods. Furthermore, if the energies $E(\gamma,t,a)$ and $E(\gamma,t,b)$, from our thought experiments, would both achieve the fastest drop to $E_{th}$ for the same $\gamma$ and at the same time, then both methods would give the same optimal damping coefficient, i.e. in that case we would have exactly $\tilde{\gamma}_{\opt}=\gamma_{\opt}^{\star}$. We see in Fig.\ \ref{fig:1Dsettime} that $\tilde{\gamma}_{\opt}\approx\gamma_{\opt}^{\star}$ for $E_{th}\rightarrow 0$. These results suggest that, in case $E_{th}\rightarrow 0$, energy $E(\gamma,t,\theta)$, i.e. \eqref{Et2}, drops the fastest to $E_{th}$ for most angles $\theta$, for approximately same damping coefficients and at approximately same times. 

\begin{figure}[h!t!]
\begin{center}
\includegraphics[width=0.48\textwidth]{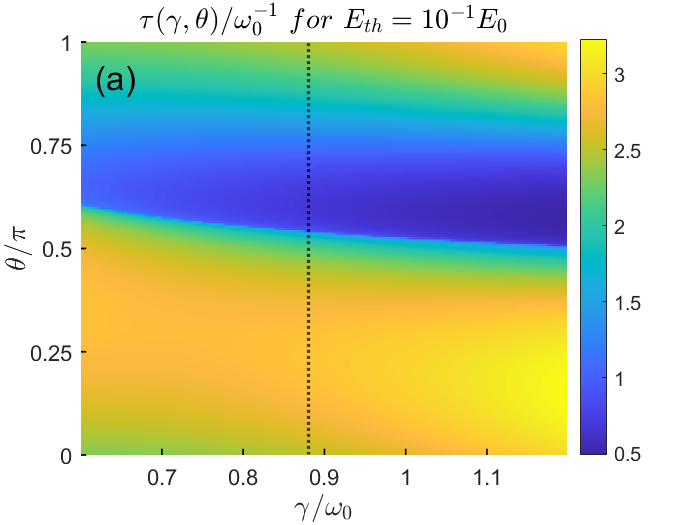}
\includegraphics[width=0.48\textwidth]{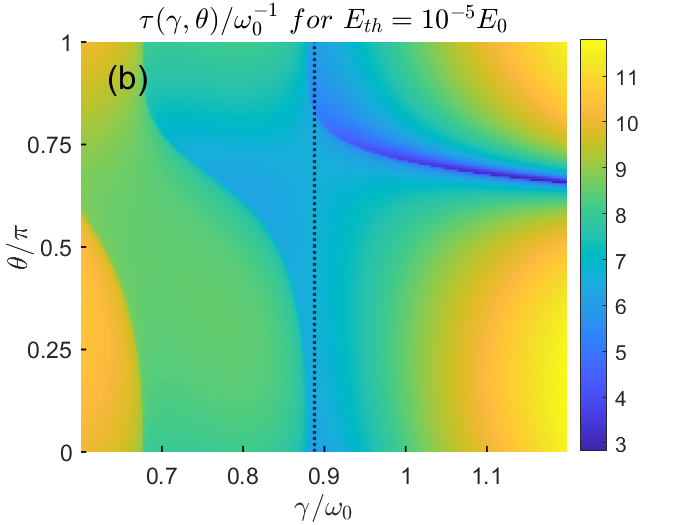}
\end{center}
\caption{Settling time of the SDOF system, i.e. $\tau(\gamma,\theta)$, for (a) $E_{th}=10^{-1}E_0$ and (b) $E_{th}=10^{-5}E_0$. On both figures, black dotted vertical lines are positioned at $\gamma_{\opt}^{\star}$ for initial conditions $\theta\in A$ and corresponding $E_{th}$.}% The step between the displayed points, along the $\gamma$-axis, is $\Delta\gamma=0.01\omega_0$.}
\label{fig:1Dminus}
\end{figure}

In Fig.\ \ref{fig:1Dminus} we see that this is indeed the case. Minima of the average settling time, for $\theta\in A$ and $E_{th}/E_0=\lbrace 10^{-1},10^{-5}\rbrace$, correspond to integrations along black dotted vertical lines shown in Fig.\ \ref{fig:1Dminus}(a) and (b). The dark blue stripe in the upper right corner of Fig.\ \ref{fig:1Dminus}(b) corresponds to initial conditions with kinetic energy greater than the potential energy and with a different signs of initial velocity and initial displacement. It is known that for this type of initial conditions the energy drops the fastest for specific damping coefficients in the overdamped regime \cite{Lelas,Lelas2}. For most of other initial conditions, the fastest drop to $E_{th}=10^{-5}E_0$ is for damping coefficients in the vicinity of the dotted vertical line in Fig.\ \ref{fig:1Dminus} (b). By looking at the areas of Fig.\ \ref{fig:1Dminus}(a) and (b) with $0\leq\theta\leq\pi/2$ and with $\pi/2\leq\theta\leq\pi$ separately, one can qualitatively understand the behaviour of $\gamma_{\opt}^{\star}$ obtained by averaging settling time over sets $B$ and $C$ as well.            

\subsection{Average settling time of the MDOF system with MPD}

\begin{figure}[h!t!]
\begin{center}
\includegraphics[width=0.45\textwidth]{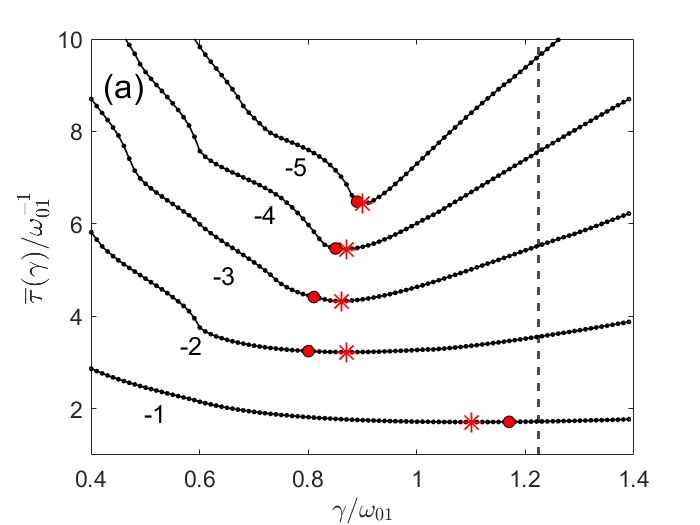}
\includegraphics[width=0.45\textwidth]{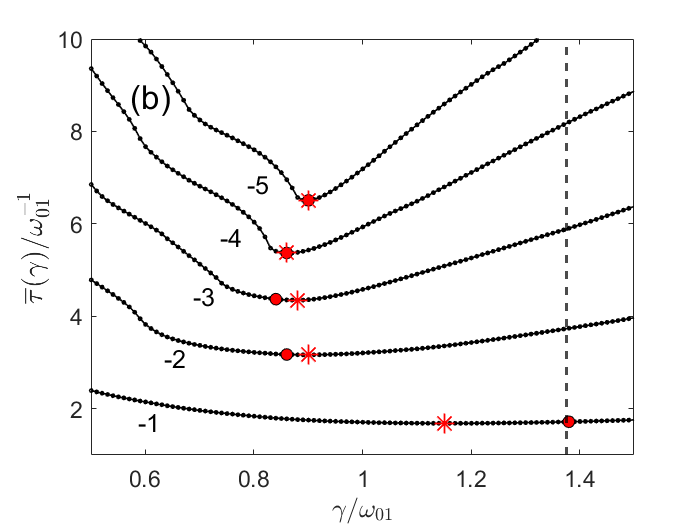}
\end{center}
\caption{Average settling time (a) of the 2-DOF system, i.e. \eqref{tauav2}, and (b) of the 3-DOF system, i.e. \eqref{tauav3}, both for $E_{0i}\in D$ $\forall i$ and $\theta_i=\theta\in A$ $\forall i$ (see text for details). The numbers $-\delta$, placed next to each curve, indicate the considered threshold $E_{th}=10^{-\delta}E_0$. The red asterisks denote points $\left(\gamma_{\opt}^{\star},\overline{\tau}(\gamma_{\opt}^{\star})\right)$, i.e. points with minimum average settling time. Filled red circles denote points $\left(\tilde{\gamma}_{\opt},\overline{\tau}(\tilde{\gamma}_{\opt})\right)$. The black dashed vertical lines are positioned at $\gamma_{\opt}$.} %The step between the displayed points, along the $\gamma$-axis, is $\Delta\gamma=0.01\omega_{01}$.}
\label{fig:23Dsettime}
\end{figure}

The settling time of the MDOF system, denoted as $\tau(\gamma,\lbrace E_{0i}\rbrace,\lbrace\theta_i\rbrace)$, is the time required for the energy of the system \eqref{EipolarTOT} to drop to a given threshold value, i.e. $E(\gamma,t,\lbrace E_{0i}\rbrace,\lbrace\theta_i\rbrace)=E_{th}$ for $t=\tau(\gamma,\lbrace E_{0i}\rbrace,\lbrace\theta_i\rbrace)$. Thus, it depends on the distribution of the initial energy $E_0$ among the modes, and on polar angles $\theta_i$, i.e. on the distribution of initial kinetic and initial potential energy within the modes. We consider all initial conditions with the same ratio of initial potential to initial kinetic energy in every mode, i.e. with $\theta_i=\theta$ $\forall i$ and $\theta\in[0,2\pi]$, and since $E_{0N}=E_0-\sum_{i=1}^{N-1}E_{0i}$, we get 
\begin{equation}
\overline{\tau}(\gamma)=\frac{1}{2\pi E_0}\int_0^{E_0}dE_{01}\int_{0}^{2\pi}d\theta\,\tau(\gamma,E_{01},\theta)\,
\label{tauav2}
\end{equation}
for the average settling time of the 2-DOF system, and
\begin{equation}
\overline{\tau}(\gamma)=\frac{1}{\pi E_0^2}\int_0^{E_0}dE_{01}\int_0^{E_0-E_{01}}dE_{02}\int_{0}^{2\pi}d\theta\,\tau(\gamma,E_{01},E_{02},\theta)\,,
\label{tauav3}
\end{equation}
for the average settling time of the 3-DOF system. We calculate \eqref{tauav2} and \eqref{tauav3} numerically, with $\theta\in A$ (see \eqref{sets}), and with
\begin{equation}
E_{0i}\in D=\lbrace \frac{jE_0}{10}\,|\,j\in\mathbb{Z},0\leq j \leq 10\rbrace\,, \forall i.
\label{setE}
\end{equation}
Thus, $dE_{01}$ and $dE_{02}$ are replaced with $\Delta E_{01}=0.1E_0$ and $\Delta E_{02}=0.1E_0$. The results are shown in Fig.\ \ref{fig:23Dsettime}. 

\begin{figure}[h!t!]
\begin{center}
\includegraphics[width=0.55\textwidth]{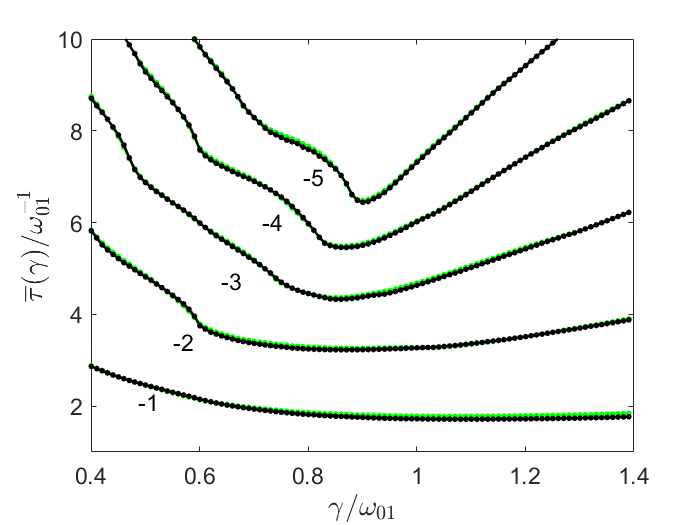}
\end{center}
\caption{Black curve shows average settling time \eqref{tauav2} calculated with $E_{01}\in D$ and $\theta\in F$. Green curve shows average settling time \eqref{tauav22} calculated with $E_{01}\in D$ and $\theta_i\in F$ for $i=1,2$. Green curve is only partially visible due to overlap with the black curve.} %The step between the displayed points, along the $\gamma$-axis, is $\Delta\gamma=0.01\omega_{01}$.}
\label{fig:2DsettimeAll}
\end{figure}

If, in addition to the initial conditions considered in relation \eqref{tauav2}, we also consider all possible initial conditions with different ratios of initial potential and initial kinetic energy within the modes, we get 
\begin{equation}
\overline{\tau}(\gamma)=\frac{1}{4\pi^2E_0}\int_0^{E_0}dE_{01}\int_{0}^{2\pi}d\theta_1\int_{0}^{2\pi}d\theta_2\,\tau(\gamma,E_{01},\theta_1,\theta_2)\,
\label{tauav22}
\end{equation}
for the average settling time of the 2-DOF system. %, and
%
%\begin{equation}
%\overline{\tau}(\gamma)=\frac{2}{E_0^2(2\pi)^3}\int_0^{E_0}dE_{01}\int_0^{E_0-E_{01}}dE_{02}\int_{0}^{2\pi}d\theta_1\int_{0}^{2\pi}d\theta_2\int_{0}^{2\pi}d\theta_3\,\tau(\gamma,E_{01},E_{02},\theta_1,\theta_2,\theta_3)\,,
%\label{tauav33}
%\end{equation}
%
%for the average settling time of the 3-DOF system.
In general, relations \eqref{tauav2} and \eqref{tauav22} will give different results for the average settling time, while, as we noted earlier, both the average energy \eqref{EipolarTOTAV} and the average energy integral \eqref{Int12av2D} are exactly the same for these two choices of sets of initial conditions. We calculate numerically \eqref{tauav22} with $E_{01}\in D$ and $\theta_i\in F=\lbrace k\pi/16\,|\,k\in\mathbb{Z},0\leq k \leq 15\rbrace$ for $i=1,2$. For comparison, we calculate again \eqref{tauav2} but with $\theta\in F$. The results are shown in Fig.\ \ref{fig:2DsettimeAll}, and we can see that \eqref{tauav2} and \eqref{tauav22} give approximately the same average settling time for the 2-DOF system we have chosen as an example, i.e. the black curves almost completely overlap with the green curves. 
%In addition, we can see that...tu sam mislio pisat o tome da uzimanje manjeg broja pocetnih uvjeta, konkretno 11*16, daje gotovo identican rezultat kao 11*360 pocetnih uvjeta, bitno je samo da su ekvidistantni na kruznici itd...ali necu o tome, osim ako recezenti se ne uhvate toga...    

%

\section{2-DOF system with non-proportional damping: minimization of the average energy integral, fastest drop of the average energy, and minimal average settling time}
\label{NonProp}

Here we consider the system as the one shown schematically in Fig.\ \ref{fig:skicaND} with $N=2$, i.e. with two masses. We take $m_1=m_2=m$, $k_1=k_2=k_3=k$, and that $c_1\geq0$ and $c_2\geq0$ can have arbitrary values. %$c_1=0.2m\omega_0$, where $\omega_0=\sqrt{k/m}$, while $c_2>0$ can have any value.
The corresponding equations of motion are
\begin{equation}
\begin{split}
m\ddot x_1(t)=-c_1\dot x_1(t)-kx_1(t)-k\left(x_1(t)-x_2(t)\right)\,,
\\m\ddot x_2(t)=-c_2\dot x_2(t)-kx_2(t)+k\left(x_1(t)-x_2(t)\right)\,.
\label{eq2D1}
\end{split}
\end{equation}
By adding (and subtracting) equations \eqref{eq2D1} and using the modal coordinates of the corresponding undamped system, i.e. %(i.e. the system with $c_1=c_2=0$)
\begin{equation}
\begin{split}
    q_1(t)=\sqrt{\frac{m}{4}}\left(x_1(t)+x_2(t)\right)
    \\q_2(t)=\sqrt{\frac{m}{4}}\left(x_1(t)-x_2(t)\right)\,,
    \label{normcoord}
\end{split}
\end{equation}
we can rewrite the system of equations \eqref{eq2D1} as
\begin{equation}
\begin{split}
\ddot q_1(t)+(\gamma_1+\gamma_2)\dot q_1(t)+(\gamma_1-\gamma_2)\dot q_2(t)+\omega_{01}^2q_1(t)=0\,
\\\ddot q_2(t)+(\gamma_1-\gamma_2)\dot q_1(t)+(\gamma_1+\gamma_2)\dot q_2(t)+\omega_{02}^2q_2(t)=0\,,
\label{eq2D2}
\end{split}
\end{equation}
where $\gamma_1=c_1/(2m)$, $\gamma_2=c_2/(2m)$, while $\omega_{01}=\omega_0$ and $\omega_{02}=\sqrt{3}\omega_0$ are the natural angular frequencies of the corresponding undamped system. We note here that, due to our choice of normalization factors in \eqref{normcoord}, the energy of the system can be written in accordance with \eqref{energyND}, i.e. we have        
\begin{equation}
E(t)=\sum_{i=1}^2\left(\frac{m\dot{x}_i(t)^2}{2}+\frac{kx_i(t)^2}{2}\right)+\frac{k(x_1(t)-x_2(t))^2}{2}=\sum_{i=1}^2\left(\dot q_i(t)^2+\omega_{0i}^2q_i(t)^2\right)\,.
\label{Ephy2D}
\end{equation}
%
%Thus, regardless of the values of $\gamma_1$ and $\gamma_2$, all possible initial conditions with the same initial energy $E_0$ can be expressed using \eqref{normcoord} at $t=0$, i.e. using the corresponding polar coordinates \eqref{polarN}.    

The defining characteristic of proportional damping is that the equations of motion decouple in the modal space (modal coordinates) of the corresponding undamped system, while in the case of non-proportional damping the equations of motion remain coupled even when transformed in the modal space of the undamped system \cite{LAZARO2016, ALVARO2021}. The system of equations \eqref{eq2D2} decouples only for $\gamma_1=\gamma_2$ ($c_1=c_2$), i.e. that case corresponds to examples with MPD studied via modal analysis in previous sections. For $\gamma_1\neq\gamma_2$ ($c_1\neq c_2$) equations \eqref{eq2D2} are coupled, i.e. the damping is non-proportional, and we can not perform modal analysis of the dynamics as in previous sections. In this case, we can proceed by transforming the system of second-order differential equations to the system of first-order differential equations, i.e. we can analyze the dynamics of the system in the state space \cite{Ves90, NakicPhd, LAZARO2016, ALVARO2021}. In order to do so, we first rewrite \eqref{eq2D2} in matrix form, i.e. as    
\begin{equation}
\ddot{\textbf{q}}(t)+\boldsymbol{\Gamma}\dot{\textbf{q}}(t)+\boldsymbol{\Omega}^2\textbf{q}(t)=0\,,
\label{matrix1}
\end{equation}
where
\begin{equation}
\textbf{q}(t)=\begin{bmatrix}
q_1(t)\\
q_2(t)
\end{bmatrix}\,\,,\,\,
\boldsymbol{\Gamma}=\begin{bmatrix}
\gamma_1+\gamma_2 & \gamma_1-\gamma_2\\
\gamma_1-\gamma_2 & \gamma_1+\gamma_2 
\end{bmatrix}\,\,,\,\,
\boldsymbol{\Omega}=\begin{bmatrix}
\omega_{01} & 0 \\
0 & \omega_{02} 
\end{bmatrix}\,.
\label{matrix2}
\end{equation}
We define the state vector
\begin{equation}
\textbf{y}(t)=\begin{bmatrix}
\boldsymbol{\Omega}\textbf{q}(t)\\
\dot{\textbf{q}}(t)
\end{bmatrix}=[\omega_{01}q_1(t)\,\,\,\omega_{02}q_2(t)\,\,\,\dot{q}_1(t)\,\,\,\dot{q}_2(t)]^{\text{T}}\,,
\label{matrix3}
\end{equation}
where superscript $\text{T}$ denotes matrix transposition. Using 
\begin{equation}
\dot{\textbf{y}}(t)=\begin{bmatrix}
\boldsymbol{\Omega}\dot{\textbf{q}}(t)\\
\ddot{\textbf{q}}(t)
\end{bmatrix}\,
\label{matrix4}
\end{equation}
and \eqref{matrix1} it is easy to show that 
\begin{equation}
\dot{\textbf{y}}(t)=\textbf{A}\textbf{y}(t)
\label{matrix5}
\end{equation}
is valid, where
\begin{equation}
\textbf{A}=\begin{bmatrix}
\boldsymbol{0} & \boldsymbol{\Omega}\\
-\boldsymbol{\Omega} & -\boldsymbol{\Gamma}
\end{bmatrix}\,\,\text{and}\,\,\boldsymbol{0}=\begin{bmatrix}
0 & 0\\
0 & 0
\end{bmatrix}\,.
\label{matrix6}
\end{equation}
The solution of \eqref{matrix5} is
\begin{equation}
\textbf{y}(t)=\exp(\textbf{A}t)\textbf{y}_0\,,
\label{matrix7}
\end{equation}
where we use notation $\textbf{y}(0)=\textbf{y}_0$. Thus, the energy \eqref{Ephy2D} can be written as
\begin{equation}
E(t)=\textbf{y}(t)^{\text{T}}\textbf{y}(t)=\textbf{y}_0^{\text{T}}\exp(\textbf{A}^{\text{T}}t)\exp(\textbf{A}t)\textbf{y}_0\,.
\label{matrix8}
\end{equation}
All possible $\textbf{y}_0$ that correspond to the same initial energy $E_0$ can be expressed using the polar coordinates \eqref{polarN}, i.e. %Thus, regardless of the values of $\gamma_1$ and $\gamma_2$, all possible initial conditions with the same initial energy $E_0$ can be expressed using \eqref{normcoord} at $t=0$, i.e. using the corresponding polar coordinates \eqref{polarN}. 
\begin{equation}
\textbf{y}_0=[\sqrt{E_{01}}\cos\theta_1\,\,\,\sqrt{E_0-E_{01}}\cos\theta_2\,\,\,\sqrt{E_{01}}\sin\theta_1\,\,\,\sqrt{E_0-E_{01}}\sin\theta_2]^{\text{T}}\,,
\label{matrix33}
\end{equation}
where we used $E_{02}=E_0-E_{01}$. Thus, the average energy, for the set of all initial conditions with initial energy $E_0$, is given by
\begin{equation}
\overline{E}(t)=\frac{1}{4\pi^2E_0}\int_0^{E_0}dE_{01}\int_0^{2\pi}d\theta_1\int_0^{2\pi}d\theta_2\,\textbf{y}_0^{\text{T}}\exp(\textbf{A}^{\text{T}}t)\exp(\textbf{A}t)\textbf{y}_0\,.
\label{matrix9}
\end{equation}
Since $$\textbf{y}_0^{\text{T}}\exp(\textbf{A}^{\text{T}}t)\exp(\textbf{A}t)\textbf{y}_0=\sum_{i=1}^4\sum_{j=1}^4(\textbf{y}_0)_i\left(\exp(\textbf{A}^{\text{T}}t)\exp(\textbf{A}t)\right)_{ij}(\textbf{y}_0)_j\,,$$ %and $$\frac{1}{E_0(2\pi)^2}\int_0^{E_0}dE_{01}\int_0^{2\pi}d\theta_1\int_0^{2\pi}d\theta_2(\textbf{y}_0)_i(\textbf{y}_0)_j=\frac{E_0}{4}\delta_{ij}\,,$$ 
we can write \eqref{matrix9} as
\begin{equation}
\overline{E}(t)=\frac{1}{4\pi^2E_0}\sum_{i=1}^4\sum_{j=1}^4\left(\int_0^{E_0}dE_{01}\int_0^{2\pi}d\theta_1\int_0^{2\pi}d\theta_2\,(\textbf{y}_0)_i(\textbf{y}_0)_j\right)\left(\exp(\textbf{A}^{\text{T}}t)\exp(\textbf{A}t)\right)_{ij}\,.
\label{matrix999}
\end{equation}
It is straightforward to calculate that the integral in expression \eqref{matrix999} vanishes for $i\neq j$, while it is equal to $(E_0\pi)^2$ for $i=j$.
%$$\int_0^{E_0}E_{01}dE_{01}=\int_0^{E_0}(E_0-E_{01})dE_{01}=\frac{E_0^2}{2}\,,$$ $$\int_0^{2\pi}\cos^2\theta_id\theta_i=\int_0^{2\pi}\sin^2\theta_id\theta_i=\pi\,,$$ $$\int_0^{2\pi}\sin\theta_i\cos\theta_id\theta_i=\int_0^{2\pi}\cos\theta_id\theta_i=\int_0^{2\pi}\sin\theta_id\theta_i=0\,,$$ 
%where 
%
%\begin{equation} \label{signum}
    %$$\delta_{ij}=\begin{cases}
%\begin{tabular}{@{}cl@{}}
   %$1$\, & if\, $i$ $=$ $j$ \\
    %$0$\, & if\, $i$ $\neq$ $j$\,,    
%\end{tabular}
    %\end{cases}$$
    %\end{equation}
%
Thus, the average energy can be written as
\begin{equation}
\overline{E}(t)=\frac{E_0}{4}\sum_{i=1}^4\left(\exp(\textbf{A}^{\text{T}}t)\exp(\textbf{A}t)\right)_{ii}=
\frac{E_0}{4}\,\text{Tr}\left(\exp(\textbf{A}^{\text{T}}t)\exp(\textbf{A}t)\right)\,,
\label{matrix99}
\end{equation}
where $\text{Tr}(...)$ denotes the trace of a matrix. The corresponding average energy integral is given by
\begin{equation}
\overline{I}=\int_0^{\infty}\overline{E}(t)dt=\frac{E_0}{4}\,\text{Tr}\left(\int_0^{\infty}\exp(\textbf{A}^{\text{T}}t)\exp(\textbf{A}t)dt\right)\,.%=\frac{E_0}{4}\,\text{Tr}\left(\textbf{X}\right)
\label{matrix10}
\end{equation}
Matrix $\textbf{X}=\int_0^{\infty}\exp(\textbf{A}^{\text{T}}t)\exp(\textbf{A}t)dt$ is the solution of the Lyapunov equation \cite{Ves90,NakicPhd}
\begin{equation}
\textbf{A}^{\text{T}}\textbf{X}+\textbf{X}\textbf{A}=-\textbf{I}\,,
\label{Lyap}
\end{equation}
where $\textbf{I}$ denotes the $4\times4$ identity matrix. Thus, we can write the average energy integral \eqref{matrix10} as
\begin{equation}
\overline{I}=\frac{E_0}{4}\,\text{Tr}\left(\textbf{X}\right)\,,
\label{matrix11}
\end{equation}
where $\textbf{X}$ is the solution of \eqref{Lyap}. Using the expressions \eqref{matrix8}, \eqref{matrix99} and \eqref{matrix11}, we can analyze the optimal damping of the 2-DOF system with non-proportional damping in a similar fashion as we did for systems with MPD in previous sections. %We note here that matrix exponentials and solutions of the Lyapunov equation, needed in \eqref{matrix8}, \eqref{matrix99} and \eqref{matrix11}, can be routinely calculated using, e.g., MATLAB, as we have done in what follows.   

\begin{figure}[h!t!]
\begin{center}
\includegraphics[width=0.48\textwidth]{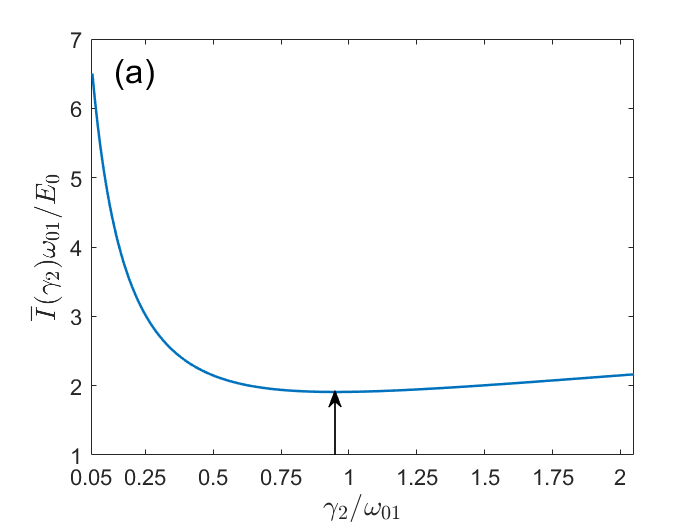}
\includegraphics[width=0.48\textwidth]{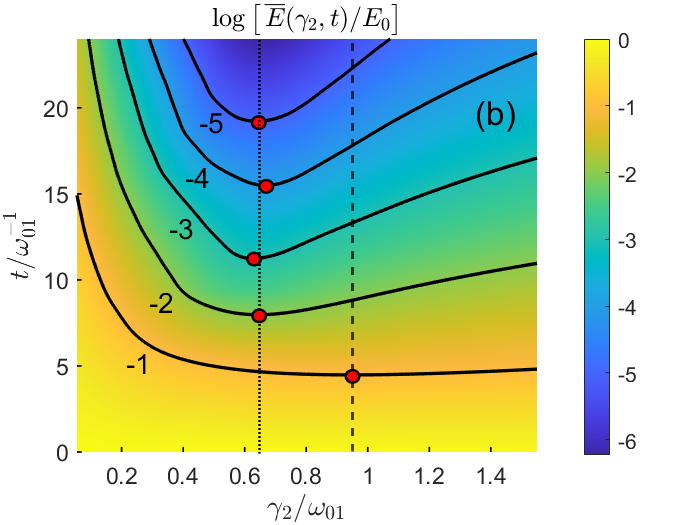}
\end{center}
\caption{(a) The average energy integral \eqref{matrix11} as a function of $\gamma_2$ for $\gamma_1=0.1\omega_{01}$. (b) The base $10$ logarithm of the average energy \eqref{matrix99} to initial energy ratio as a function of $\gamma_2$ and $t$, i.e. $\log\left(\,\overline{E}(\gamma_2,t)/E_0\right)$, for $\gamma_1=0.1\omega_{01}$. See text for details.}% The step between the displayed points, along the $\gamma$-axis, is $\Delta\gamma=0.01\omega_0$.}
\label{fig:dodatak1}
\end{figure}

As an example, we consider a system described by equations \eqref{eq2D2} with a fixed value $\gamma_1=0.1\omega_{01}$ and analyze the optimal value of $\gamma_2$ adapted to the set of all initial conditions with initial energy $E_0$. In Fig.\ \ref{fig:dodatak1}(a), we show the average energy integral \eqref{matrix11} as a function of $\gamma_2$ for $\gamma_1=0.1\omega_{01}$. We note here that we again use $\omega_{01}^{-1}$ as the time unit, thus, the average energy integral is in $E_0/\omega_{01}$ units. The minimum of the average energy integral is obtained for $\gamma_{\opt}=0.95\omega_{01}$ (rounded to two decimals), as indicated by the arrow in Fig.\ \ref{fig:dodatak1}(a). %Thus, minimization of the average energy integral gives $\gamma_{\opt}=0.95\omega_{01}$ as the optimal value of $\gamma_2$.
In Fig.\ \ref{fig:dodatak1}(b), we show the base $10$ logarithm of the average energy \eqref{matrix99} to initial energy ratio as a function of $\gamma_2$ and $t$, i.e. $\log\left(\,\overline{E}(\gamma_2,t)/E_0\right)$, for $\gamma_1=0.1\omega_{01}$. The contour lines denote points with $\overline{E}(\gamma_2,t)/E_{0}=\lbrace 10^{-1}, 10^{-2},10^{-3},10^{-4},10^{-5}\rbrace$ respectively, as indicated by the numbers placed to the left of each contour line. Furthermore, the filled red circles in Fig.\ \ref{fig:dodatak1}(b) denote the points on the contour lines that are closest to the $\gamma_2$ axis, i.e. corresponding to the optimal damping coefficient $\tilde{\gamma}_{\opt}$ given by the fastest drop of the average energy, the black dashed vertical line is positioned at the optimal damping coefficient $\gamma_{\opt}=0.95\omega_{01}$, and the black dotted vertical line is positioned at $\gamma_2=0.65\omega_{01}$. We can see that $\tilde{\gamma}_{\opt}$ overlaps with $\gamma_{\opt}$ only for the energy threshold $E_{th}=10^{-1}E_0$, while $\gamma_{\opt}$ has about $45\%$ higher value than $\tilde{\gamma}_{\opt}$ for $E_{th}/E_0=\lbrace 10^{-2},10^{-3},10^{-4},10^{-5}\rbrace$, since $\tilde{\gamma}_{\opt}\approx0.65\omega_{01}$ for these values of $E_{th}$. Thus, for the fixed value $c_1=0.2m\omega_{01}$ in \eqref{eq2D1}, the minimization of the average energy integral gives $c_2=1.90m\omega_{01}$ as optimal, and the fastest drop of the average energy gives approximately the same $c_2$ as optimal, but only for $E_{th}=10^{-1}E_0$, while it gives $c_2\approx1.30m\omega_{01}$ as optimal for $E_{th}/E_0=\lbrace 10^{-2},10^{-3},10^{-4},10^{-5}\rbrace$.       

\begin{figure}[h!t!]
\begin{center}
\includegraphics[width=0.48\textwidth]{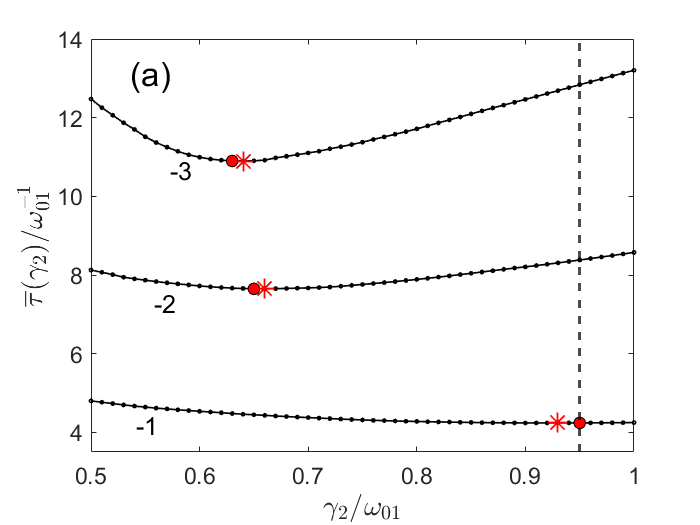}
\includegraphics[width=0.48\textwidth]{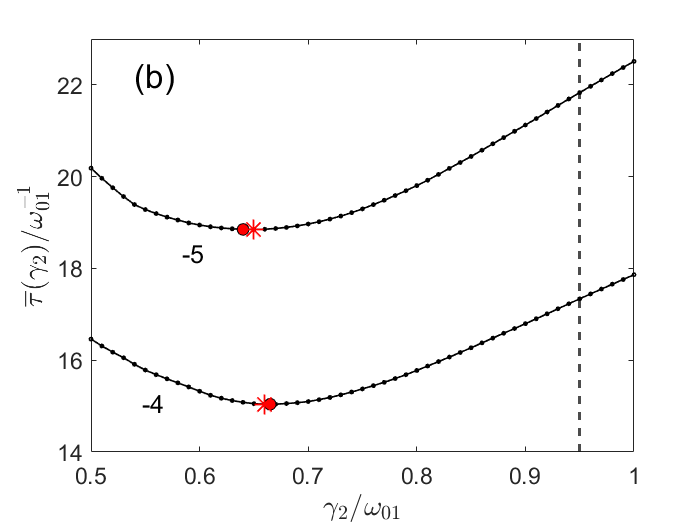}
\end{center}
\caption{Average settling time \eqref{tauav2NP} for (a) $E_{th}/E_0=\lbrace10^{-1},10^{-2},10^{-3}\rbrace$ and (b) $E_{th}/E_0=\lbrace10^{-4},10^{-5}\rbrace$. The numbers $-\delta$, placed next to each curve, indicate the considered threshold $E_{th}=10^{-\delta}E_0$. The red asterisks denote points $\left(\gamma_{\opt}^{\star},\overline{\tau}(\gamma_{\opt}^{\star})\right)$, i.e. points with minimum average settling time. Filled red circles denote points $\left(\tilde{\gamma}_{\opt},\overline{\tau}(\tilde{\gamma}_{\opt})\right)$. The black dashed vertical lines are positioned at $\gamma_{\opt}$. See text for details.}% The step between the displayed points, along the $\gamma$-axis, is $\Delta\gamma=0.01\omega_0$.}
\label{fig:dodatak2}
\end{figure}

The settling time of the system described by the equations \eqref{eq2D2} with fixed value $\gamma_1=0.1\omega_{01}$, denoted $\tau(\gamma_2,\textbf{y}_0)$, is the time required for the corresponding energy \eqref{matrix8} to drop to a given threshold value $E_{th}$. Thus, it depends on $\gamma_2$ and the initial state vector \eqref{matrix33}. The average settling time, with respect to all possible $\textbf{y}_0$ that correspond to the same initial energy $E_0$, is given by  
\begin{equation}
\overline{\tau}(\gamma_2)=\frac{1}{4\pi^2E_0}\int_0^{E_0}dE_{01}\int_{0}^{2\pi}d\theta_1\int_{0}^{2\pi}d\theta_2\,\tau(\gamma_2,\textbf{y}_0)\,,
\label{tauav2NP}
\end{equation}
and the value of $\gamma_2$ that minimizes \eqref{tauav2NP} will be denoted by $\gamma_{\opt}^{\star}$. We calculate \eqref{tauav2NP} numerically for the discrete set of initial conditions with $E_{01}\in\lbrace jE_0/10\,|\,j\in\mathbb{Z},0\leq j \leq 10\rbrace$ and $\theta_i\in\lbrace j\pi/8\,|\,j\in\mathbb{Z},0\leq j \leq 15\rbrace$ for $i=1,2$. The results are shown in Fig.\ \ref{fig:dodatak2}(a) and (b). We point out the coordinates of the minimal average settling time (red asterisks), and the values of the average settling time obtained for optimal values of $\gamma_2$ given by the other two methods, i.e. for $\tilde{\gamma}_{\opt}$ (filled red circle) and $\gamma_{\opt}$ (dashed vertical line), for comparison. We can see that $\tilde{\gamma}_{\opt}\approx\gamma_{\opt}^{\star}$ for all chosen values of $E_{th}$, while $\gamma_{\opt}\approx\gamma_{\opt}^{\star}$ only for $E_{th}=10^{-1}E_0$.    
%

%We note here that matrix exponentials and solutions of the Lyapunov equation can be routinely calculated using, e.g., MATLAB, as we have done in what follows.

\section{Discussion}

We can ask the following question. If we randomly excite the free vibrations of the SDOF or MDOF system with MPD over a long period of time, so that each time one initial condition from some given set of initial conditions is realized, and that each new vibration begins only after the previous one has settled, which of the three presented optimal damping coefficients (i.e. $\gamma_{\opt}$, $\tilde{\gamma}_{\opt}$ or $\gamma_{\opt}^{\star}$) would be the best choice in the long run if we want the system to spend as little time as possible vibrating with energy greater than $E_{th}$? It is clear, by definition, that the optimal damping coefficient $\gamma_{\opt}^{\star}$, which gives the minimal average settling time, is the best choice in that case. Furthermore, if $E_{th}$ corresponds to the measurement resolution of the apparatus with which we observe the system, then for $\gamma_{\opt}^{\star}$ the system spends the least amount of time (effectively) out of equilibrium. We showed that the optimal damping coefficient $\tilde{\gamma}_{\opt}$, which gives the fastest drop of the average energy to $E_{th}$, is an excellent approximation of $\gamma_{\opt}^{\star}$ (the better the smaller values of $E_{th}$ are considered), while the optimal damping coefficient $\gamma_{\opt}$, which minimizes the average energy integral, can be significantly different than $\gamma_{\opt}^{\star}$, depending on the set of initial conditions and $E_{th}$ considered.          

Of course, with an increase in the number of degrees of freedom and an increase in the range of expected initial conditions, it becomes more and more computationally demanding to determine the optimal damping coefficients $\gamma_{\opt}^{\star}$, since we need to determine the settling time separately for each initial condition in order to calculate the average settling time for the considered set of initial conditions. In case it is impractical to determine $\gamma_{\opt}^{\star}$, $\tilde{\gamma}_{\opt}$ can serve as the second best choice. It is much easier to determine $\tilde{\gamma}_{\opt}$ because only the average energy needs to be analyzed, i.e. a single quantity that includes all initial conditions from the considered set of initial conditions and which, at least in case of MPD, is easily obtained analytically.  

In the case of a system with damping that does not allow analytical treatment, energy and average energy, as functions of time and magnitudes of individual dampers, can be determined numerically by studying the vibrating system as a first-order ordinary differential equation with matrix coefficients, as we did in Section \ref{NonProp}. Thus, we can numerically analyze the time evolution of energy and of average energy, with respect to some finite discrete set of initial conditions, and find a set of damping parameters for which the average settling time has a minimum and a set of damping parameters for which the average energy drops to a desired energy threshold the fastest. Of course, this approach can be applied only for systems with a moderate number of degrees of freedom and a small number of dampers, due to the rapid growth of the parameter space that needs to be searched. Furthermore, the number of initial conditions that we consider must be in accordance with the computational possibilities.

The two new methods we propose for determining the optimal damping adapted to sets of initial conditions can also be applied to the types of damping that we did not study here. For example, in the case of a system with Rayleigh damping \cite{katsikadelis2020dynamic}, both the energy for a specific initial condition and the average energy with respect to some set of initial conditions can be determined analytically using modal analysis, and based on that analytical expression, it can be numerically investigated for which values of the mass and stiffness proportionality constants the average settling time has a minimum and for which values of the mass and stiffness proportionality constants the average energy drops the fastest to some energy threshold. A similar approach could also be used in the case of Caughey damping \cite{Caughey1,Caughey2}, where the diagonalizable $n$-DOF system has $n$ damping parameters.  
 
%Despite limitations in the size of the system that can be treated , the two new methods for determining overall optimal damping with respect to a given set of initial conditions should be investigated in more detail in the context of vibrating systems with damping that does not allow modal analysis because, as we have shown, they give results that are significantly different from the results given by the already known method. This will be a topic of our future work.   

\section{Conclusion and outlook}

%For a single and multi degree of freedom systems, we determined the optimal damping coefficients adapted to different sets of initial conditions using the known method of minimizing the (zero to infinity) time integral of the energy of the system, averaged over a set of initial conditions, and using two new methods that we introduced, i.e. . 
We introduced two new methods for determining the optimal damping of free vibrations adapted to different sets of initial conditions. One method is based on determining the damping for which the energy of the system, averaged over a set of initial conditions, drops the fastest to a given threshold value. The other method is based on determining the damping that gives minimal average settling time of the system, where we take that the system settled when its energy dropped to a given threshold value. We showed that the two new methods give results for optimal damping that are in excellent agreement with each other but differ significantly from the results obtained by the well-known method of minimizing the average energy integral. More precisely, for MDOF systems with MPD and for considered sets of initial conditions, the two new methods yielded optimal damping coefficients that converge to the critical damping of the first mode as the target energy threshold decreases. On the other hand, for these same systems and sets of initial conditions, the method of minimizing the average energy integral yielded optimal damping coefficients which are deep in the overdamped regime with respect to the first mode. We also showed how to analyze systems with non-proportional damping. For the considered 2DOF system with non-proportional damping, we found that the new methods yield approximately the same values of the optimal damping coefficients for all considered energy thresholds and that these coefficients differ significantly from the optimal damping coefficient given by the minimization of the average energy integral. 

%Despite the limitations on the size of the systems, i.e. on the number of degrees of freedom and the number of dampers, for which it is computationally practical to determine the optimal damping using our two new methods, they may be applied for the analysis of some real systems. 
%Our two new methods may be applied for the analysis of some real systems. 
The two new methods presented in this paper can be applied to the analysis of optimal damping of free vibrations in real systems as long as they are modeled with a moderate number of degrees of freedom and we are interested in the optimization with respect to a small number of dampers. For example, optimal damping of free vibrations is studied in the context of the reduction of residual vibrations in ultra-precision manufacturing machines which are modeled as 2-DOF systems \cite{Residual1,Residual2}. The quarter car (2-DOF) and half car (4-DOF) models are used to analyze the optimal damping of the car suspension \cite{CarModel}. The action of a tuned mass damper is often analyzed considering a 2-DOF system, i.e. a primary SDOF system equipped with a tuned mass damper \cite{gutierrez2013tuned}. We have listed here only a few examples in which simple models of practically important real systems are considered, and for which our two methods for determining optimal damping of free vibrations can be straightforwardly applied. We have shown that two new methods give significantly different results compared to one of the already known methods, i.e. to minimization of the average energy integral, and it is worth investigating whether similar differences exist in comparison to other already known methods, such as, e.g., the optimization of the $H_2$ and $H_{\infty}$ norms of the frequency response functions which are used to analyze the optimal damping of residual vibrations in \cite{Residual1,Residual2}. We envision that the application of the two new methods to a detailed analysis of the optimal damping of free vibrations in the mentioned (and similar) real systems will result in new interesting insights important from both theoretical and practical perspectives.
This will be the topic of our next work.

%In these works, the $H_2$ and $H_{\infty}$ norms of the frequency response functions are used to analyze the optimal damping of residual vibrations. It would be interesting to compare the results obtained by our methods with the results obtained in 2.

\section{Acknowledgments}

The author is grateful to Ivica Nakić for useful discussions. This work was supported by the QuantiXLie Center of Excellence, a project co-financed by the Croatian Government and European Union through the European Regional Development Fund, the Competitiveness and Cohesion Operational Programme (Grant No. KK.01.1.1.01.0004).
\\
The author has no conflicts to disclose.

\bibliographystyle{elsarticle-num}  
\bibliography{aipsamp2}

\end{document}